\documentclass[a4paper,11pt]{article}

\usepackage{jheppub} 

\usepackage[T1]{fontenc} 

\def\eps{\varepsilon}

\def\sig{\sigma} \def\kap{\kappa} \def\lam{\lambda}

\def \RR {{\mathbb R}}
\def \CC {{\mathbb C}}
\def \ZZ {{\mathbb Z}}
\def \NN {{\mathbb N}}

\def \TT {{\mathcal T}}
\def\Eps{{\mathcal E}}
\def \eins {{\mathbb I}}
\def \norm#1{{\vert\!\vert #1\vert\!\vert}}

\def \merw#1#2{M^{#2}_{#1}}
\def \wdp#1{d\mu_{#1}(p)}

\def\lra{\leftrightarrow}
\def\lrd#1{\stackrel{\leftrightarrow}{\partial_{#1}}}

\def\tpf{two-point function}
\def\tpk{two-point kernel}
\def\pll{Pauli-Lubanski limit}
\def\set{stress-energy tensor}
\def\sloc{string-localized}
\def\ploc{point-localized}

\def\bpm{\begin{pmatrix}} \def\epm{\end{pmatrix}}

\newcommand{\ul}{\underline}
 
\newcommand{\ket}[1]{\vert #1\rangle}

\def \HH {\mathcal{H}}

\def\be{\begin{equation}} \def\bea{\begin{eqnarray}} 
\def\ba{\begin{array}} \def\eea{\end{eqnarray}} 
\def\ee{\end{equation}}\def\ea{\end{array}}
\def\inv{^{-1}}
\def\wh{\widehat} \def\wt{\widetilde}
\def\ul{\underline} \def\ol{\overline}

 \def\inv{^{-1}}

\def\bpm{\begin{pmatrix}} \def\epm{\end{pmatrix}}

\newcommand{\wick}[1]{{:}#1{:}} 
\def\Stab{\mathrm{Stab}} \def\SO{\mathrm{SO}} \def\so{\mathrm{so}}
\newcounter{allcount} \numberwithin{allcount}{section}

\newtheorem{defn}[allcount]{Definition}
\newtheorem{lemma}[allcount]{Lemma}
\newtheorem{propo}[allcount]{Proposition}
\newtheorem{coro}[allcount]{Corollary}
\newtheorem{remark}[allcount]{Remark}
\newtheorem{example}[allcount]{Example}
\newtheorem{conj}[allcount]{Conjecture}

\newcommand{\dref}[1]{Def.~\ref{#1}}
\newcommand{\eref}[1]{Eq.~(\ref{#1})}

\newcommand{\cref}[1]{Cor.~\ref{#1}}
\newcommand{\lref}[1]{Lemma~\ref{#1}}
\newcommand{\pref}[1]{Prop.~\ref{#1}}
\newcommand{\rref}[1]{Remark~\ref{#1}}
\newcommand{\xref}[1]{Example~\ref{#1}}
\newcommand{\sref}[1]{Sect.~\ref{#1}}
\newcommand{\aref}[1]{App.~\ref{#1}}

\newcommand{\cjref}[1]{Conj.~\ref{#1}}

\def\qed{\hfill$\square$\medskip}

\title{Pauli-Lubanski limit and stress-energy tensor \\
for infinite-spin fields}
\author[a]{Karl-Henning Rehren}
\affiliation[a]{Institute for Theoretical Physics,
  Georg-August-University G\"ottingen,\\ 
  Friedrich-Hund-Platz 1, 37077 G\"ottingen, Germany}
\emailAdd{rehren@therie.physik.uni-goettingen.de}

\abstract{String-localized quantum fields transforming in Wigner's
  infinite-spin representations were originally introduced in
  \cite{MSY1,MSY2}. We construct these fields as limits of fields of
  finite mass $m\to0$ and finite spin $s\to\infty$. We determine a \sloc\
  infinite-spin quantum \set\ with a novel prescription that does not
  refer to a classical Lagrangean.  
}


\dedicated{Dedicated to Klaus Fredenhagen on the occasion of his 70th birthday}

\begin{document}

\maketitle

\flushbottom

\section{Introduction}
\label{s:intro}
\setcounter{equation}{0}

\subsection{Quantum fields in the infinite-spin representations}
Starting from the posit that one-particles states look like
one-particle states in every inertial frame, Wigner concluded that
particles should be identified with unitary positive-energy
representations of the (proper orthochronous) Poincar\'e group or its
twofold covering. His famous classification \cite{W39} contains massive 
spin representations, massless helicity representations, and two
one-parameter families of true resp.\ projective massless
representations called ``infinite spin'' or ``continuous spin''. The
parameter $\kappa^2>0$ is the eigenvalue of the Pauli-Lubanski
operator $W^2 = (P\wedge M)^2$ ($P_\rho$ is the momentum and
$M_{\sig\tau}$ are the Lorentz generators). 

Weinberg \cite{W} showed how one can associate local (or anti-local,
in the projective case according to the spin-statistics theorem)
quantum fields to all these representations except infinite spin. Let
us from now on consider only the true (bosonic) representations. 

In the massless case, only the field strengths can be constructed as
local fields on the Fock space over the sum of Wigner representations
with helicities $\pm h$ (which is irreducible when the parity is
included). Their potentials necessarily violate
either locality and covariance (e.g., in the Coulomb gauge) \cite{W}, 
or they must be constructed on an indefinite-metric Krein space of
which the Fock space over the Wigner representation is a quotient. The
latter option underlies all gauge-theoretic approaches to modern
quantum field theory, with its introduction of more unphysical ghost
fields to ``compensate'' states of negative probability.

For the irreducible infinite-spin representations, Yngvason \cite{Y}
has shown a no-go theorem, that covariant local Wightman fields cannot
exist. This was taken for a long time as a serious reason to consider
these representations as ``unphysical''.  

This conclusion is, however, a misunderstanding. From work of
Buchholz and Fredenhagen \cite{BF} we know that quantum operators 
connecting scattering states with the vacuum may not in general be 
assumed to be localized in bounded spacetime regions. Instead, the
best localization that can be proven is (a narrow ``spacelike cone''
arising by smearing) a ``string'' $S_e(x)=\{x+se: s\geq 0\}$ ($e^2=-1$). 
So string-localization of interpolating fields may be a necessary feature
of charged states in interacting theories, and only observables need
to be \ploc. While the result in \cite{BF} applies to massive
theories, it has a massless counterpart in the violation of Lorentz
covariance of charged sectors of QED \cite{FMS}, due to the presence of
``photon clouds'' attached to charged fields that can -- because of
Gau\ss' law -- at best be localized in a narrow cone.

String-localization surprisingly emerged in the context of
infinite-spin representations, when Mund, Schroer and Yngvason
\cite{MSY1}, based on work by Brunetti, Guido and Longo 
\cite{BGL}, discovered a construction of fields $\varphi(e,x)$
transforming in the infinite-spin representation, that
are localized along strings. Two such fields commute with each other
whenever their strings $S_e(x)$ and $S_{e'}(x')$ are spacelike separated.  

The same authors also noticed that string-localized free fields may also be
useful in the case of finite spin: E.g., one can construct a potential
$A_\mu(e,x)$ of the Maxwell field strength $F_{\mu\nu}(x)$ directly on
the ghost-free Fock space of the latter. This potential is manifestly
presented as a string-integral over the field strength: 
\bea\label{Imax}
A_\mu(e,x) := \int_0^\infty ds\, F_{\mu\nu}(x+se)e^\nu .
\eea
Similar constructions are possible for any mass and finite spin
(\cite{PY} and \cite{MRS1} for the massless case, \eref{ass} below and
\cite{MRS1} for the massive case). Such potentials have the benefit
\cite{S15,S16} that 
\begin{enumerate}\item[(i)] in the
massive case: they have an improved short-distance behaviour compared
to the point-localized Proca potentials, which is a promising feature
when the free fields are used perturbatively to set up an interacting
theory; and unlike the latter, they admit a massless limit;
\item[(ii)] in the massless case: they are directly defined on the
  physical Hilbert space without the need to introduce Gupta-Bleuler
  conditions or compensating ghost fields. 
\end{enumerate}
These potentials were systematically studied for every integer spin
$s$ in \cite{PY}, and in \cite{MRS1,MRS2} with the focus on their
massless limit at fixed $s$, and on the
discrepancy between the massless string-localized potentials and the
massless limit of the massive string-localized potentials. 

In contrast to the string-localized fields of finite spin like \eref{Imax}, 
the infinite-spin fields of \cite{MSY1,MSY2} are 
``intrinsically string-localized'': they cannot be expressed as
string-integrals over \ploc\ fields. This makes them quite non-trivial
objects to study.  

Also from different points of view, there is a lot of renewed interest
in the infinite-spin representations. Schuster and Toro \cite{ST1,ST2} 
and Rivelles \cite{R} study quantum wave equations in a
one-particle setting. Also their wave functions depend on an auxiliary
four-vector $e$ which has, however, no direct geometric
interpretation. ``Localization'' in the sense of causal commutators
has no meaning in a one-particle (quantum-mechanical) approach. In
\cite{ST3}, Schuster and Toro write down canonical commutation
relations which are local both in $x$ and $e$, and it is not clear to
the author how they realize such commutation relations in a Hilbert
space -- which would be at variance with Yngvason's no-go theorem. 

Bekaert et al.\ \cite{BM,BMN,BS} are pursuing a ``Fierz-Pauli
program'' attempting to identify a classical action principle leading to
wave equations compatible with the infinite-spin representation. 
With quantization beyond the scope of this program, the constraints
due to Hilbert space positivity and causal commutation relations
(addressed in, e.g., \cite{S17,MRS1}) play no role in their work.  

In contrast, our work is placed in the setting of ``Wigner quantization'',
where free fields are associated with a unitary Wigner
representation by 
\bea\label{fields}
\phi_M(x) = \int d\mu_0(p) \sum\nolimits_n\big[u_{Mn}(p)(\vec
k)a_n^*(p) \, e^{ipx} + v_{Mn}(p)a_n(p,\vec k) \, e^{-ipx} \big], 
\eea
and the matrices $u$ and $v$ are ``intertwiners'' (see \eref{itw-gen}) 
between the unitary Wigner representation of the Lorentz group and the
matrix representation (typically a tensor product of Lorentz matrices)
under which the field transforms. They are needed to absorb the 
``Wigner rotations'' that would otherwise spoil the covariant
transformation law. For hermitean Bose fields, $v(p)=\ol{u(p)}$. 

In this setting, Hilbert space positivity is manifest from the outset
because \eref{fields} is defined on the Fock space over the Wigner
representation. Field equations, \tpf s and commutation relations
follow without the need of a variational principle and ``canonical''
equal time commutation relations, i.e., they follow intrinsically
(except for the choice of the localizing intertwiner functions) from
Wigner representation theory. In this setting, the non-existence of a
covariant quantum Maxwell potential on the physical Hilbert space is
just the fact that the interwining relation has no solution. 

This approach offers also an important new flexibility in perturbation
theory \cite{S17}. While the Wigner representation fixes all
properties of the {\em particles}, the choice of intertwiners
determines (among other things) the short-distance behaviour of free
{\em fields} that create these particles from the vacuum. Since UV 
singularities are a major problem of
perturbation theory, one can benefit from the fact that   
\sloc\ fields have better UV properties and therefore admit
renormalizable couplings that do not exist with \ploc\ fields (if the
latter exist at all on a ghost-free Hilbert space). The case of
massive vector bosons may be taken as an example: in order to control
the UV problems of \ploc\ massive vector fields, the prevalent
prescription treats them as massless fields in an indefinite-metric
Hilbert space which ``behave as if they were massive'' thanks to the
Higgs mechanism. In the \sloc\ approach, one may instead start with
massive vector bosons and interpolating fields in their physical
Hilbert space from the outset. (The Higgs field is still needed, but
for a different reason, see below.)

In all cases of interactions mediated by \sloc\ fields, one has to
observe that the obvious hazard of violating causality through the use of
\sloc\ Lagrangeans in perturbation theory, can be controlled in terms of
a certain cohomological ``pair condition'' \cite{S17} on the
interaction terms: It secures the string-independence of the classical
action, and is the first order condition for the string-independence
of the perturbative quantum causal S-matrix. Higher order conditions may
require additional ``induced'' interaction terms. 

E.g., in the
presence of selfinteracting massive vectormesons one needs an
additional coupling of vectormesons to an Hermitian (Higgs) field in
oder to uphold second order renormalizability. (Such compensations
between fields with different spins have hitherto been expected to
take place in the presence of supersymmetry; but whereas
there are serious problems to maintain supersymmetry in second order,
in the case of selfinteracting massive vector mesons this compensation
is the very raison d'\^etre for the Higgs particle -- without
invoking a mechanism of spontaneous symmetry breaking, and without the
need of unphysical ghost degrees of freedom.) 

In the case of infinite-spin particles, the use of \sloc\ fields is
not a choice (to improve the UV behaviour) but an intrinsic
necessity \cite{BGL}. Whether a pair condition can be fulfilled for any
interaction with ordinary particles, is presently unknown. Schroer
\cite{S17} discusses indications why this might not be the case (for
infinite spin or already for some finite spin beyond a maximal
value). As a consequence, these particles would be invisible in 
detectors (``inert''); the identification of a stress-energy tensor in the
present work may be a starting point in order to investigate whether they
might at least cause semiclassical gravitational back reactions.  

The reader only interested in the infinite-spin \set, may jump
directly to \sref{s:sets}, retaining from the preceeding sections only 
the properties \pref{p:Ur} of the \sloc\ fields. These
properties, although derived through the \pll\ of finite mass and
finite spin, refer directly to the Fock space over the
infinite-spin representation, so that the construction of the \set\ is
intrinsic.

\subsection{Contents and plan of the paper}

{\bf \pll.} We study in \sref{s:PLL}, how the string-localized infinite-spin
fields of \cite{MSY1} are approximated by massive string-localized
fields of finite spin, in the ``\pll'' $s\to\infty$ at fixed
Pauli-Lubanski parameter $\kappa^2=m^2\cdot s(s+1)$. 
This limit is suggested by the fact that the Pauli-Lubanski operator
$W^2 = (P\wedge M)^2$ is a Casimir operator of the Lie algebra of the
Poincar\'e group with eigenvalue $m^2\cdot s(s+1)$ in the
representation $(m,s)$, and with eigenvalue $\kappa^2$ in the massless
infinite-spin representation $U^1_\kappa$. 

The \pll\ is well known for the Wigner representations themselves \cite{McK},
basically because the massless little group $E(2)$ \cite{W39} is a
contraction of the massive little group $SO(3)$, see \aref{a:ind}. In
the infinite-spin representations, the pseudo-translations 
(the subgroup $\RR^2\subset E(2)$ embedded into the proper
orthochronous Lorentz group
$SO(1,3)_+^\uparrow$) are non-trivially represented with spectrum of their
generators lying on a circle of radius $\kappa$. The corresponding
basis of eigenfunctions $e^{im\varphi}$ of the rotations of the little
group ($m\in\ZZ$ is the magnetic quantum number) can be approximated
by the eigenfunctions $Y_{lm}$ ($l=s\to\infty$, $-l\leq m\leq l$) of the
finite-spin representations of $SO(3)$, see \aref{a:ind}.

But a ``lift'' of the limit of representations to the associated quantum 
fields is not known so far. An obvious obstruction seems to be that the
(conserved and traceless) Proca potentials have a number of indices 
increasing with $s$, so that they are not even candidates for a
``converging family of fields''. Another obvious obstruction is that a
limit of local commutator functions, if it exists, should be a local,
and not a string-local commutator function that we know to be the best
possible thing for infinite spin. But the most important obstruction
is the singularity of the Proca potentials at $m=0$ which become
stronger with increasing spin. It is related to the non-existence of
point-localized currents and \set s for the massless 
representations of finite helicity \cite{WW}. 

With string-localized massive potentials, these obstructions are
absent. The potentials are manifestly string-localized from the
beginning, they are regular at $m=0$, and they are neither traceless
nor conserved, so that one may consider their divergences (called
``escort fields'') of fixed rank as natural candidates for converging
families. Indeed this will turn out to be true, see \sref{s:PLL}.

This finding is a bit surprising. In \cite{MRS1}, we have found that
the ``scalar escort field'' converges in the massless limit at fixed
$s$ to a true massless scalar field, while we are now claiming that in
the \pll, it converges to an infinite-spin field!
Indeed, there is no contradiction. At finite mass, the scalar escort
is coupled to the other escorts of any rank $r\leq s$ by field
equations (\eref{desc}, \eref{asc}), and each escort carries the
entire spin $s$ representation. This coupling goes to zero in
the massless limit at fixed $s$. But as our results implicitly show,
it ``remains stable'' when $s$ increases at the same time, so that the
limit field carries the entire infinite-spin representation. 

\bigskip

{\bf Stress-energy tensor.} In \cite{MRS1}, we have constructed
currents and \set s for the massive finite spin representations that
have a regular massless limit at fixed $s$. In the second part of our
paper (\sref{s:sets}), we present a general construction that produces
string-localized such densities also for the infinite-spin 
representations, and elucidate whether these exist as Wightman fields. 
In fact, this is expected not to be the case: their vacuum
\tpf s are expected to diverge due to the infinitely many inner degrees of
freedom that are summed over, and we give indications that this is
indeed the case. While the vacuum \tpf\ is tedious to
compute exactly, it is very easy to 
compute the thermal one-point function in KMS states. Here, the
expected divergence \cite{W48} proportional to $2s+1$ (in accord with
the thermodynamical equipartition theorem) can be explicitly seen.

On the other hand, the commutator of the densities with the fields is
a derivation that integrates to the infinitesimal gauge or Poincar\'e
transformations. Because the latter are meaningful also at infinite
spin, we expect the limit of the densities to exist at least ``as
derivations'' on the algebra of fields. 

Studying the existence and properties of currents and \set s for the
infinite-spin representations is of great interest, because even a
classical Langrangean from which these could possibly be derived, is
not known (see the ``Fierz-Pauli program'' of \cite{BS}). 
The intricacies of the quantum field theory, due to the conflict
between Hilbert space positivity and causal point-localization, can 
only be overcome with string-localized fields. 

In \sref{s:infspin}, we review the essential features of \sloc\
infinite-spin fields and introduce the special fields that will appear
in the \set. \sref{s:preps} prepares the ground
for the \pll. After these preparations, the initial main result of
\sref{s:PLL}, \pref{p:a-lim}, which entails everything else, is very
quickly obtained.

\section{String-localized infinite-spin fields}
\label{s:infspin}
\setcounter{equation}{0}

The authors of \cite{MSY1,MSY2} constructed string-localized fields
$\phi(e,x)$ on the Fock space over the infinite-spin
representation $U^1_\kappa$ that transform like
\bea\label{cov}
U_\kappa(a,\Lambda)\phi(e,x)U_\kappa(a,\Lambda)^* = \phi(\Lambda e,\Lambda
x+a).
\eea
The action of $U^1_\kappa(a,\Lambda)$ on the one-particle space is
constructed, as in Wigner's original approach \cite{W39}, by induction
from a representation $d_\kappa$ of the stabilizer group $E(2)$ of the
reference four-momentum $p_0=(1,0,0,1)^t$, and a family of Lorentz
transformations $B_p$ for every $p\in H_0=\{p\in\RR^4:p^2=0,p^0>0\}$, 
such that $B_pp_0=p$. The representation space $\HH_\kappa=L^2(\kappa
S^1)$ of $d_\kappa$ are the square-integrable functions of a
two-dimensional vector $\vec k$, $\vec k^2=\kappa^2$. On such
functions, the rotations and pseudo-translations act like 
\bea
\label{Dkap}
(d_\kappa(R_\alpha) f)(\vec k) = f(R_{-\alpha}\vec k), \qquad
(d_\kappa(T_{\vec a}) f)(\vec k)=e^{i\vec a\cdot\vec k}f(\vec k). 
\eea

The irreducible one-particle representation $U^1_\kappa$ induced from
$d_\kappa$ is defined on square-integrable functions on the zero mass
shell $H_0$ with values in $\HH_\kappa$, see \aref{a:ind}. It
immediately lifts to the representation $U_\kappa$ on the Fock space.
The construction of hermitean fields out of creation and annihilation
operators $a(p,\vec k)$, $a^*(p,\vec k)$ ($p\in H_0$, $\vec k\in \kappa S^1$)
then proceeds in terms of ``intertwiners'' $u(e,p)$: 
\bea
\label{infspin}
\phi_u(e,x) = \int d\mu_0(p) \int
d\mu_\kappa(\vec k) \big[u(e,p)(\vec k)a^*(p,\vec k) \, e^{ipx} +
\ol{u(e,p)(\vec k)}a(p,\vec k) \, e^{-ipx} \big] \quad
\eea
The intertwiners are distributions in $p$ and $e$ with values in
$\HH_\kappa$, that satisfy
\bea\label{itw}
u(\Lambda e,\Lambda p) = d_\kappa(W_{\Lambda,p})u(e,p) \qquad
(\Lambda\in \SO(1,3)_+^\uparrow),\eea
where $W_{\Lambda,p}=B_{\Lambda p}\inv \Lambda B_p\in E(2)$ is the
Wigner ``rotation''\footnote{Of course, it is not a rotation, nor is
  $B_p$. 
We sloppily adopt the terminology Wigner ``rotation'' and standard 
``boost'' for $B_p:p_0\mapsto p$ from the massive case.}.
This property ensures the transformation law \eref{cov}. 
In order to ensure that the commutator function vanishes for spacelike  
separated strings, it is crucial \cite[Thm.~3.3]{MSY1}
that $u(e,p)$ is analytic in the $e$ variable in the complex tube 
$\TT_+=\{e\in\CC^4: e^2=-1, \mathrm{Im}\,e \in V_+\}$, and satisfies
certain local bounds in the tube as specified in
\cite[Def.~3.1]{MSY1}. The analyticity is necessary to ensure locality
by a contour deformation argument. The bounds ensure that the
boundary value of the analytic function $u$ at $\mathrm{Im}\,e\to0$ 
defines an operator-valued distribution.

The intertwiner condition \eref{itw} is equivalent to the pair of
relations
\bea\label{itw1}
u(e,p) = u(B_p\inv e,p_0) \qquad (p\in H_0)
\eea
(which determines $u(e,p)$ for all $p\in H_0$ as soon as
$u_0(e)=u(e,p_0)$ is given), and
\bea
\label{itw2}
u_0(We)= d_\kappa(W)u_0(e) \qquad (W\in E(2)),
\eea
which determines $u_0(e)$ along each orbit of $e$ under the little
group. The argument is standard: Let $\xi_0=\frac12(1,0,0,-1)^t$ and
$\xi(\vec a):=T_{\vec a}\xi_0=\frac12(\vec a^2+1,2a^1,2a^2,\vec a^2-1)^t$.
The orbits of $E(2)$ are parametrized by the invariants $e^2$ and
$(ep_0)$, and $e$ along each orbit with $(ep_0)\neq0$ is parametrized by 
$e(\vec a)=\frac{e^2}{2(ep_0)}p_0 + (ep_0)\xi(\vec a)$, $\vec a\in \RR^2$. 
Because $e_0=e(\vec 0)$ is fixed by the rotations of $E(2)$,
$u_0(e_0)(\vec k)$ is a constant function of $\vec k\in \kappa
S^1$ by the first of \eref{Dkap}. Its value $f(e^2,(ep_0))$ is an
arbitrary function of the orbit. The translations of $E(2)$ then determine 
$$u_0(e)(\vec k) = f(e^2,(ep_0))\cdot e^{-i\frac{(eE(\vec
    k))}{(ep_0)}}$$
everywhere along the orbit, by the second of \eref{Dkap}. Here,
$E:\RR^2\to\RR^4$ is the standard embedding $E(\vec k):=
(0,k^1,k^2,0)^t$ into Minkowski space. The function $f(x,y)$ is not
determined by \eref{itw2}. By \eref{itw1}, one gets 
\bea
\label{uf} 
u_f(e,p)(\vec k) = f(e^2,(ep))\cdot 
e^{-i \frac{(eE_p(\vec k))}{(ep)}}
\eea
where $E_p(\vec k):=B_pE(\vec k)$. 

These are identical with the ``smooth'' solutions to the three differential
equations (3.6)--(3.8) in \cite{ST1}, that are the 
infinitesimal version of \eref{itw2}, lifted to $p\in H_0$ by \eref{itw1}. 
The ``singular'' solutions supported on the orbits with $(ep)=0$ are
not admissible as intertwiners. 

\subsection{The scalar standard field}

Analyticity in the forward tube $\TT_+$ of $e$ requires to take
$1/(ep)$ in \eref{uf} as the distribution $1/(ep)_+$, because 
$(ep)\in\CC_+$ if $e\in \TT_+$. The simplest solution 
\bea\label{ut0}
\wt u^\kappa(e,p)(\vec k) := e^{-i \frac{(eE_p(\vec k))}{(ep)_+}}
\eea
solves the wave equations Eq.~(3.17)--(3.18) in \cite{ST1} along with
their subsidiary conditions Eq.~(3.19)--(3.20) with $n=0$. But it does
not satisfy the bounds required to prove the vanishing of commutators
when the strings are spacelike separated \cite[Thm.~3.3]{MRS1}. 
E.g., for $p=(\mu,0,0,\mu)^t$ and $e=(ix,\sqrt{1-x^2},0,0)^t\in
\TT_+$, the estimate $\norm{\wt u^\kappa(e,p)}\geq 
\exp\big(\frac{\sqrt{1-x^2}}{2x}\frac\kappa\mu\big)$ violates every 
power law bound as $x\to0$ and $\mu\to0$. Instead:
\begin{propo}\label{p:k-bd} 
(see also \cite[pp.~42ff]{K} and Remark after \cite[Def.~3.1]{MSY1})
In the tube $\TT_+$, it holds 
$\vert e^{-i \frac{(eE_p(\vec k))}{(ep)}}\vert \leq \vert e^{\frac{i\kappa}{(ep)}}\vert$. 
Therefore, the solution 
\bea\label{u-std} u^{\kappa(0)}(e,p)(\vec k) =
e^{-i\kappa\frac{\sqrt{-e^2}}{(ep)_+}}\cdot 
e^{-i \frac{(eE_p(\vec k))}{(ep)_+}}
\eea
at real $e$ exists as a weakly continuous $L^2(\kappa S^1)$-valued
function. It is the intertwiner of a \sloc\ field $\phi^{\kappa(0)}(e,x)$.     
\end{propo}
\begin{remark}\label{homo}
Although we always put $e^2=-1$, we write $\sqrt{-e^2}$ because we are
going to take derivatives w.r.t.\ $e$ by defining intertwiners as
homogeneous functions $u(\lambda e):= u(e)$ ($\lambda>0$) for all
spacelike $e$. In the sequel, we shall refer to \eref{u-std} as the
``standard intertwiner'', and the associated field $\phi^{\kappa(0)}$
as the ``standard (\sloc\ infinite-spin) field''. We call  
$\omega^\kappa(e,p)=e^{-i\kappa\frac{\sqrt{-e^2}}{(ep)_+}}$ the ``K\"ohler
factor'' \cite{K}. It appears only in the combination \eref{u-std}
where it cancels the essential singularity of \eref{ut0}.
\end{remark}
Proof of \pref{p:k-bd}: If $e=e'+ie''$ is in the tube, then $e''\in V_+$ and
$(e'e'')=0$, hence $e'$ is spacelike (or $=0$). $e^2=-1$ implies
$-1<e'^2\leq 0$ and $0<e''^2\leq 1$.  

For $e\in\TT_+$, the real part of the exponent 
$-i\frac{(eE_p(\vec k))}{(ep)}$ is 
$$\frac{(e''E_p(\vec k))(e'p)-(e'E_p(\vec k))(e''p)}{\vert
  (ep)\vert^2} = \frac{(f''E(\vec k))(f'p_0)-(f'E(\vec
  k))(f''p_0)}{\vert (fp_0)\vert^2}$$ 
where $f:=B_p\inv e = f'+if''$,
$(f'f'')=0$, $f''\in V_+$, $f'^2-f''^2=-1$. Parametrize
$f'=\alpha'p_0+\beta'\xi_0+E(\vec f')$,
$f''=\alpha''p_0+\beta''\xi_0+E(\vec f'')$, such that 
$(\vec f'\vec f'')=\alpha'\beta''+\alpha''\beta'$, and
$\beta''>0$, $\vec f''^2<2\alpha''\beta''$, and
$2\alpha'\beta'-\vec f'^2-2\alpha''\beta''+\vec f''^2=-1$, hence 
$\vec f'^2<2\alpha'\beta'+1$.

Then, the numerator is (cf.\ \cite{Gz}) 
\bea\notag
(\beta''\vec f'-\beta'\vec f'')\cdot\vec k\leq\kappa\vert\beta''\vec
f'-\beta'\vec f''\vert = \kappa\sqrt{\beta''^2\vec f'^2+\beta'^2\vec
  f''^2-2\beta'\beta''(\vec f'\vec f'')}\leq\\\notag
\leq\kappa\sqrt{\beta''^2(2\alpha'\beta'+1)+\beta'^22\alpha''\beta''-2\beta'\beta''(\alpha'\beta''+\alpha''\beta')}=\kappa\beta''.
\eea
Thus $\vert e^{-i\frac{(eE_p(\vec k))}{(ep)}}\vert\leq
e^{\kappa\frac{\beta''}{\vert(ep_0)^2}} = \vert \omega^\kappa(e,p)\inv\vert$.
It follows that $\omega^\kappa(e,p)\cdot e^{-i \frac{(eE_p(\vec
    k))}{(ep)}}$ is an analytic function in $\TT_+$ bounded by 1. 
Thus it satisfies the bounds specified in
\cite[Def.~3.1]{MRS1}
and the remark following it, hence its boundary value \eref{u-std} is
well-defined as a function, and defines a string-localized field
by \cite[Thm.~3.3]{MRS1}.  \qed

Mund, Schroer and Yngvason in \cite{MSY2} also gave solutions to the
intertwiner relation \eref{itw} in a different form (for $e^2=-1$): 
\bea\label{UF} 
U_F(e,p)(\vec k) = \int d^2a \, e^{i\vec k\cdot \vec a}
\, F\big((eB_p\xi(\vec a))\big).
\eea
The function $F(z)$ must be analytic and polynomially bounded in the upper
half-plane, hence its Fourier transform $\wh F$ is supported on
$\RR_+$. One can bring this form into the form \eref{uf}:
With the Fourier representation of $F(z)$, the $\vec a$-integration
becomes Gaussian and can be performed when $e\in \TT_+$, with the
result (see \cite{Gz})
\bea
U_F(e,p)(\vec k) = u^{\kappa(0)}(e,p)(\vec k)\cdot \Big[\frac{2\pi i}{(ep)_+}
\int_0^\infty \frac{dt}t\, \wh F\Big(\frac{\kappa}{\sqrt{-e^2}}\,t\Big)\, 
e^{-i\frac{\kappa\sqrt{-e^2}}{(ep)_+}\frac{(t-1)^2}{2t}}\Big]. \quad
\eea
The choice $2\pi \wh F\big(\frac{\kappa}{\sqrt{-e^2}}\,t\big)=\delta(t-1)$ 
gives the intertwiner $U_F=\frac{i}{(ep)_+}u^{\kappa(0)}$, i.e., the
standard field is $\phi^{\kappa(0)}(e,x)=-(e\partial_x)\phi_F(e,x)$. 

We introduce the string-integration operator
\bea
\label{I}
(I_eX)(x):= \int_0^\infty ds \, X(x+se),
\eea
already occurring in \eref{Imax}. If $X(e,x)$ is localized along the
string $e$, then so is $I_eX(e,x)$. One has 
\bea\label{Iinv}
(e\partial_x)I_e X=I_e(e\partial_x)X=-X. 
\eea
In momentum space, acting on $e^{ipx}$, this is the multiplication
operator by 
\bea
\label{Ip}
I_e(p)=\frac i{(ep)_+}\equiv
\lim_{\eps\searrow 0}\frac i{(ep)+i\eps}
\eea
as a distribution. 
\newpage
We also introduce 
\bea
\label{Jx}
(J_e)_\mu{}^\nu := I_e[e^\nu\partial_\mu - \delta_\mu^\nu (e\partial_\nu)] 
= \delta_\mu^\nu + I_ee^\nu\partial_\mu, 
\eea
because of \eref{Iinv}. In momentum space, this is the multiplication
operator by 
\bea
\label{Je} 
J_e(p)_\mu{}^\nu = (ep)\partial_{e^\mu}
\frac{e^\nu}{(ep)_+} = \delta_\mu^\nu - p_\mu
\frac{e^\nu}{(ep)_+},
\eea
so that
\bea\label{deu}
(ep)\partial_{e^\mu} \,\wt u^\kappa(e,p)(\vec k) = -i
(J_e(p)E_p(\vec k))_\mu \cdot \wt u^\kappa(e,p)(\vec k).
\eea
Varying w.r.t.\ $e$, one has  
\bea
\label{deJ} 
(ep)\partial_{e^\lam} J_e(p)_\mu{}^\nu = -p_\mu J_e(p)_\lam{}^\nu. 
\eea
Because we are going to take derivatives w.r.t.\ $e$ of intertwiners
multiplied with the K\"ohler factor, it is convenient to introduce
\bea
\label{De} 
D_e(p):=\partial_e
-\frac{i\kappa}{\sqrt{-e^2}}\frac{J_e(p)e}{(ep)_+},
\eea
so that $D_e(p)(\omega^\kappa\wt u(e,p))=\omega^\kappa \partial_e \wt
u(e,p)$. Acting on the corresponding fields, this is the operator
\bea
\label{Dex}
D_e = \partial_e
-\frac{\kappa}{\sqrt{-e^2}}I_eJ_ee.
\eea
\begin{propo}\label{p:eoms} The standard string-localized infinite-spin 
  field $\phi^{\kappa(0)}(e,x)$ satisfies the equations of motion 
\bea
\label{eom1}
\square_x \phi^{\kappa(0)} =
(e\partial_e)\phi^{\kappa(0)} = (\partial_xD_e)\phi^{\kappa(0)}  =
((e\partial_x)^2D_e^2 +\kappa^2) \phi^{\kappa(0)} =0.
\eea
The last two equations in \eref{eom1} are equivalent to
\bea
\label{eom2}
\sqrt{-e^2}\,(\partial_x\partial_e)\phi^{\kappa(0)} = \sqrt{-e^2}\,
(e\partial_x)\square_e \phi^{\kappa(0)} = -\kappa\,\phi^{\kappa(0)}. 
\eea
The Pauli-Lubanski equation follows:  
\bea\label{eom3}
W^2\phi^{\kappa(0)} = \kappa^2 \phi^{\kappa(0)},
\eea
where $W^2=\square_x(e^2\square_e-(e\partial_e)^2-(e\partial_e)) 
+2(e\partial_x)(\partial_x\partial_e)(e\partial_e)
-e^2(\partial_x\partial_e)^2- (e\partial_x)^2\square_e$
is the Pauli-Lubanski operator.
\end{propo}
Proof: The Klein-Gordon equation is fulfilled by construction, and the
homogeneity in $e$ is manifest from \eref{u-std}. Using \eref{deu} and
$p^2=(pE_p(\vec k))=0$, hence $(pJ_eE_p(\vec k))=0$ and $(J_eE_p(\vec
k))^2=(E_p(\vec k))^2= -\kappa^2$, one computes 
$$(p\partial_e)\,\wt u^\kappa(e,p) = 0,\quad 
((ep)^2\square_e-\kappa^2)\,\wt u^\kappa(e,p) =0,$$ 
hence 
$$(pD_e)\,u^{\kappa(0)}(e,p) = 0, \quad 
((ep)^2D_e^2-\kappa^2)\,u^{\kappa(0)}(e,p) =0.$$
\newpage 
These relations for the intertwiner are equivalent to the last two
equations in \eref{eom1}. The first of \eref{eom2} is equivalent to
the third in \eref{eom1} by \eref{Iinv}, and the second of \eref{eom2}
follows from the last of \eref{eom1} by a lengthy calculation using
also the second and third of \eref{eom1}. 
\eref{eom3} is then a consequence of the previous. 
\qed

\subsection{Tensor fields I}
\label{s:tensor1}

We shall later also need tensor fields that transform like 
\bea
\label{cov-r}
U_\kappa(a,\Lambda)\phi_{\mu_1\dots\mu_r}(e,x)U_\kappa(a,\Lambda)^* =
\prod \Lambda_{\mu_i}^{\,\,\nu_i}\,\phi_{\nu_1\dots\nu_r}(\Lambda e,\Lambda
x+a).
\eea
They are formed with intertwiners that, regarded as functions with
values in $(\RR^4)^{\otimes r}\otimes\HH_\kappa$, satisfy 
\bea
\label{itw-r}
u(\Lambda e,\Lambda p) = \big((\Lambda^{\otimes
  r})\otimes d_\kappa(W_{\Lambda,p})\big)\, u(e,p).
\eea
To simplify notation, we write contractions as $a^\mu b_\mu\equiv
(ab)\equiv a^tb$, and for symmetric tensors of rank $r$, we write 
$$X(v) \equiv v^{\mu_1}\dots v^{\mu_r}X_{\mu_1\dots\mu_r} \equiv v^{t\otimes r}X
\qquad (v\in\RR^4).$$
The tensor components are recovered by differentiation w.r.t.\
$v^\mu$.
\begin{defn}
\label{d:ur-def}
We introduce the symmetric rank $r$ tensor intertwiners
\bea
\label{ur-def}
u^{\kappa(r)}(e,p)(\vec k) := (-\kappa)^{-r} 
(J_e(p)E_p(\vec k))^{\otimes r}\cdot u^{\kappa(0)}(e,p)(\vec k), 
\eea
where $E_p:\RR^2\to\RR^4$ is the anti-isometric embedding 
($E_p(\vec k)^2 = -\vec k^2$) introduced in \eref{uf}. The associated
tensor fields are called $\phi^{\kappa(r)}(e)$.  
\end{defn}
\begin{propo}
\label{p:rec} 
The symmetric tensor fields $\phi^{\kappa(r)}$ are recursively related
to $\phi^{\kappa(0)}$ by  
\bea
\label{phi-rec} 
-\kappa \phi^{\kappa(r+1)}(e,x,v) =
\big(r\cdot(v\partial_x)+(e\partial_x)(vD_e)\big)\phi^{\kappa(r)}(e,x,v).
\eea
\end{propo}
Proof: \eref{deu} and $\big(r\cdot p_\mu+(ep)\partial_{e^\mu}\big)\circ 
J_e^{\otimes r} = J_e^{\otimes r}(ep)\circ\partial_{e^\mu}$ imply 
$$-i(v^tJ_e(p)E_p(\vec k))^{r+1}\cdot \wt u^{\kappa}(e,p)(\vec k) =
\big(r\cdot(vp)+(ep)(v\partial_{e})\big)(v^tJ_e(p)E_p(\vec k))^r\cdot \wt
u^{\kappa}(e,p)(\vec k).$$
Restoring the K\"ohler factor, we have to replace $\partial_e$ by
$D_e(p)$ defined in \eref{De}:  
\bea
\label{u-rec} 
i\kappa u^{\kappa(r+1)}(e,p,v)(\vec k) = 
\big(r\cdot(vp)+(ep)(vD_e(p))\big)u^{\kappa(r)}(e,p,v)(\vec k).
\eea
This is the momentum space version of \eref{phi-rec}. \qed

We may take \eref{phi-rec} as recursive definitions. The covariance of
\eref{phi-rec} and \eref{u-rec} ensures that $\phi^{\kappa(r)}$ and
their intertwiners $u^{\kappa(r)}$ transform according to \eref{cov-r}
and \eref{itw-r}, respectively, and the bounds of
\cite[Def.~3.1]{MRS1} are inherited from the bound of $u^{\kappa(0)}$
because $(J_e(p)E_p(\vec k))^2=-\kappa^2$. 
\begin{coro}
\label{phi}
$\phi^{\kappa(r)}_{\mu_1\dots\mu_r}(e,x)$ are string-localized tensor fields. 
\end{coro}

\subsection{Tensor fields II}
\label{s:tensor2}
\begin{defn}
\label{d:Ur-def}
We introduce a second family of symmetric rank $r$ tensor fields 
\bea
\label{Phir-def}
\Phi^{\kappa(r)}(e,x) = 2^{r/2}\cdot 
\Big(
S_r\sum\nolimits_{2k\leq r}\gamma^r_k\cdot 
(J_e^{\otimes 2}\eta\circ\eta^t)^{\otimes k} \Big) \phi^{\kappa(r)}(e,x)
\eea
with the coefficients 
$\gamma^r_k=\frac{1}{4^kk!}\frac{r!}{(r-2k)!}\frac{1}{(1-r)_k}$, and 
$S_r$ the projection onto the symmetric tensors in $(\RR^4)^{\otimes r}$. 
The associated \sloc\ tensor fields are called $\Phi^{\kappa(r)}(e)$. 
$J_e$ is the operator given in \eref{Jx}.
(Trivially, $\Phi^{\kappa(0)}=\phi^{\kappa(0)}$ and
$\Phi^{\kappa(1)}=\sqrt2\cdot \phi^{\kappa(1)}$.)
\end{defn}
\begin{remark}
\label{Phi} 
The operations on $\phi^{\kappa(r)}$ that define $\Phi^{\kappa(r)}$
  preserve the localization and covariance \eref{cov-r}. Thus, the
  latter are again \sloc\ tensor fields.  
\end{remark}
The following proposition exhibits the advantage of the new fields
$\Phi^{\kappa(r)}$, that will become important in \sref{s:sets}. 
\begin{propo}
\label{p:Ur} 
{\rm (i)} The intertwiner of $\Phi^{\kappa(r)}$ equals 
\bea\label{Ur}
U^{\kappa(r)}(e,p)(\vec k) = \frac{2^{r/2}}{(-\kappa)^r}\cdot (J_e(p)
E_p)^{\otimes r} \big(\Pi_2^{(r)}(\vec k^{\otimes r})\big) \cdot
u^{\kappa(0)}(e,p)(\vec k). 
\eea
Here, $\Pi_2^{(r)}$ is the projection (of two-dimensional range) onto
the symmetric traceless tensors in $(\RR^2)^{\otimes r}$.   \\
{\rm (ii)} For $\vec k=\kappa\bpm\cos\varphi\\[-.5mm] \sin\varphi\epm$ 
and $\eps_\pm = \frac1{\sqrt2}\bpm 1\\[-.5mm]\pm i\epm$, one has the
decomposition into helicity eigenstates 
\bea\label{eigen}
\frac{2^{r/2}}{\kappa^r}\Pi_2^{(r)}(\vec k^{\otimes r}) = 
\left\{\ba{ll} 1&\quad (r=0)\\
e^{-ir\varphi}\eps_+^{\otimes r} +
e^{ir\varphi}\eps_-^{\otimes r} &\quad (r>0),\ea\right.
\eea
hence $U^{\kappa(0)}=u^{\kappa(0)}$, and for $r>0$
\bea
\label{UrEps}
U^{\kappa(r)}(e,p)(\vec k) = (-1)^r \Big(e^{-ir\varphi}
\Eps_+(e,p)^{\otimes r} + e^{ir\varphi}\Eps_-(e,p)^{\otimes r}\Big)\cdot
u^{\kappa(0)}(e,p)(\vec k),\quad
\eea
where $\Eps_\pm(e,p):=J_e(p)E_p\eps_\pm$.
\end{propo}
\begin{remark}
\label{r:sharp}
The presence of the factor $u^{\kappa(0)}(\vec k)$ prevents the
interpretation of $\Phi^{\kappa(r)}(e)$ as ``fields of sharp helicity'', 
in accord with the irreducibility of the infinite-spin representation.
\end{remark}
Proof of \pref{p:Ur}: By \eref{Phir-def}, the intertwiner of
$\Phi^{\kappa(r)}$ is  
\bea
\label{Ur-def}
U^{\kappa(r)}(e,p)(\vec k) = 2^{r/2}\cdot \Big(
S_r\sum\nolimits_{2k\leq r}\gamma^r_k\cdot 
(J_e(p)^{\otimes 2}\eta\circ\eta^t)^{\otimes k} \Big) 
u^{\kappa(r)}(e,p)(\vec k)
\eea
with $J_e(p)$ given in \eref{Je}. 
By \eref{ur-def}, the intertwiner $u^{\kappa(r)}$ is in the range of
the operator $(J_e(p)E_p)^{\otimes r}$. Therefore, we may consider
the operators 
$$ J_e(p)^{\otimes 2}\eta\circ\eta^t(J_e(p)E_p)^{\otimes
  2}:(\RR^2)^{\otimes 2}\to(\RR^4)^{\otimes 2}$$
appearing in \eref{Ur-def} when \eref{ur-def} is inserted. Using the
standard vectors $p_0=(1,0,0,1)^t$ and $\xi_0=(\frac12,0,0,-\frac12)^t$, 
we have $\eta=-E^{\otimes 2}\delta_2+p_0\otimes\xi_0+\xi_0\otimes p_0$
as tensors in $(\RR^4)^{\otimes 2}$, where 
$\delta_2=\vec e_1\otimes\vec e_1+\vec e_2\otimes\vec e_2$ as a tensor
in $(\RR^4)^{\otimes 2}$. Applying the standard ``boost'' $B_p:p_0\mapsto p$, 
we get 
$$\eta=-E^{\otimes 2}\delta_2+p\otimes B_p\xi_0+B_p\xi_0\otimes p.$$
Because $p^2=0$, we have $\eta^tJ_e(p)^{\otimes
  2}=\eta^t-(B_p\xi_0)^t\otimes p^t-p^t\otimes (B_p\xi_0)^t$, hence 
$$J_e(p)^{\otimes
  2}\eta = (J_e(p)E_p)^{\otimes
  2}\delta_2, \quad \eta^t(J_e(p)E_p)^{\otimes 2}=\eta^tE_p^{\otimes
  2}=\delta_2^tE_p^{t\otimes 2}E_p^{\otimes 2}=\delta_2^t,$$
because $p$ is orthogonal to the range of $E_p$. Thus, \eref{Ur-def}
can be rewritten as  
$$U^{\kappa(r)}(e,p)(\vec k) = \frac{2^{r/2}}{\kappa^r}\cdot 
(J_e(p)E_p)^{\otimes r} \Big(
S_r\sum\nolimits_{2k\leq r}\gamma^r_k\cdot 
(\delta_2\delta_2^t)^{\otimes k} \otimes \eins_2^{\otimes r-2k}\Big)
\vec k^{\otimes r}\cdot u^{\kappa(0)}(e,p)(\vec k).
$$
The operator in brackets is the projection $\Pi_2^{(r)}$
onto the symmetric traceless tensors in $(\RR^2)^{\otimes
  r}$ (this is in fact the defining property of the coefficients
$\gamma^r_k$ \cite{G}). This proves the claim (i). Now, write $\vec k
= \frac\kappa{\sqrt2}(e^{-i\varphi}\eps_++e^{+i\varphi}\eps_-)$. Then
\eref{eigen} is a well-known identity (that may be proven by
induction in $r$), and \eref{UrEps} follows. \qed 
\begin{propo}\label{p:Eoms}
Besides the massless Klein-Gordon equation, the infinite-spin
symmetric tensor fields $\Phi^{\kappa(r)}_{\mu_1\dots\mu_r}(e,x)$
satisfy the equations of motion and constraints 
\bea\label{eoms}\eta^{\mu\nu} \Phi^{\kappa(r)}_{\mu\nu\mu_3\dots\mu_r}
= 0,\quad \partial^\mu
\Phi^{\kappa(r)}_{\mu\mu_2\dots\mu_r}=0,\quad e^\mu
\Phi^{\kappa(r)}_{\mu\mu_2\dots\mu_r}=0,
\\[2mm]
\label{e-var}
(e\partial_e)\Phi^{\kappa(r)}_{\mu_1\dots\mu_r} = 0,\qquad
(\partial_xD_e)\Phi^{\kappa(r)}_{\mu_1\dots\mu_r} =0, \hspace{10mm}\eea
as well as the coupling relations for $r\geq2$ 
\bea\label{coup}
\Big((e\partial_x)D_{e^{\mu_{1}}}\Phi^{\kappa(r)}_{\mu_2\dots\mu_{r+1}} +
\sum\nolimits_{i=2}^{r+1}\partial_{\mu_i}
\Phi^{\kappa(r)}_{\mu_1\dots\mu_{i-1}\mu_{i+1}\dots\mu_{r+1}}\Big)
+ (\mu_1\lra\mu_{2})
= \hspace{20mm} \notag \\
=-\sqrt2\kappa \Big(\Phi^{\kappa(r+1)}_{\mu_1\dots\mu_{r+1}}
-\frac12(E_e)_{\mu_1\mu_{2}}\Phi^{\kappa(r-1)}_{\mu_3\dots\mu_{r+1}}\Big). 
\hspace{10mm}
\eea
Here, $(E_e)_{\mu\nu}$ is the integral- and differential operator 
$((J_e\otimes J_e)\eta)_{\mu\nu}$. For $r=0$, the second term in the
bracket on the r.h.s.\ is absent, and for $r=1$ is replaced by 
$-(E_e)_{\mu_1\mu_2}\Phi^{\kappa(0)}(v)$.  
\end{propo}
Proof: We proceed in momentum space, where $\partial_x=ip$ on the
intertwiners \eref{UrEps}. \eref{eoms} and \eref{e-var} follow by
a direct computation, using $(\Eps_\pm\Eps_\pm)=0$ and
$(p\Eps_\pm)=(e\Eps_\pm)=(p\partial_e)\Eps_\pm=0$, as well as
$D_eu^{\kappa(0)}=\omega^\kappa\partial_e\wt u^\kappa$. For the coupling
relations, apply $i(ep)D_{e^{\mu_{1}}}$ to $U^{\kappa(r)}_{\mu_2\dots\mu_{r+1}}$. 
Use $\partial_{e^\mu}\Eps_{\pm\nu} = -p_\nu\Eps_{\pm\mu}$
by \eref{deJ}, which cancels the spatial derivatives on the l.h.s.\ of
\eref{coup}, and $i(ep)D_{e^\mu}u^{\kappa(0)} = (J_e(p)E_p(\vec k))_\mu 
u^{\kappa(0)} = \frac\kappa{\sqrt2}(e^{-i\varphi}\Eps_++e^{i\varphi}\Eps_-)
u^{\kappa(0)}$ by \eref{deu}. This produces the term 
$(\mathrm{id}+\pi_{12})(\wt\Eps_++\wt\Eps_-)\otimes
\Big(\wt\Eps_+^{\otimes r}+\wt\Eps_-^{\otimes r}\Big)$ where
$\wt\Eps_\pm\equiv e^{\mp i\varphi}\Eps_\pm$ and $\pi_{12}$ is the
permutation of the first two tensor factors. This tensor trivially
equals 
$$=2\Big(\wt\Eps_+^{\otimes r+1}+\wt\Eps_-^{\otimes r+1}\Big) + 
(\Eps_+\otimes\Eps_- + \Eps_-\otimes\Eps_+)\otimes
\Big(\wt\Eps_+^{\otimes r-1}+\wt\Eps_-^{\otimes r-1}\Big).$$
Finally, $\Eps_+\otimes\Eps_- +
\Eps_-\otimes\Eps_+=(J_e(p)E_p)^{\otimes 2}(\delta_2)=- E_e(p)$, as in
the proof of \pref{p:Ur}, implies the claim. \qed

\section{Preparations for the \pll}
\label{s:preps}
\setcounter{equation}{0}

\subsection{Two-point functions and commutators: basics}
\label{s:basics}
Our fields are free fields, so that the entire information resides
in their \tpf s, which in particular determine the commutator functions. 
It is convenient to express the \tpf s as integral
or differential operators acting on the canonical scalar \tpf\
$(\Omega,\varphi(x)\varphi(y)\Omega) = \Delta_m(x-y)$. This amounts to
the insertion of a ``\tpk'' into the Fourier representation: 
\bea
\label{MXY}
(\Omega,X(x)Y(y)\Omega) = \int \wdp{m}\,e^{-ip(x-y)}\cdot \merw m{X,Y}(p),
\eea
where $\wdp{m} = (2\pi)^{-3}d^4p\,\delta(p^2-m^2)\theta(p^0)$ is the
Lorentz invariant measure on the positive energy mass shell $H_m$.
E.g., the canonical scalar field and the Proca field have the \tpk s
\bea
\label{PP}
\merw{m}{\varphi,\varphi}=1\quad\hbox{resp.}\quad
\merw{m}{A^{\rm P}_\mu,A^{\rm P}_\nu}(p)
= -\pi_{\mu\nu}\equiv -(\eta_{\mu\nu} -\frac{p_\mu p_\nu}{m^2}).
\eea

\subsection{Two-point functions of string-localized spin 1 fields}
For the massive Proca field we introduce the \sloc\ potential by the same
formula as for the Maxwell potential \eref{Imax}; but in this case it
can also be expressed in terms of the \ploc\ potential $A_\nu^{\rm P} =
-\frac 1{m^2}\partial^\mu F^{\rm P}_{\mu\nu}$ that exists on the Hilbert
space: 
\bea
\label{Iproca}
A_\mu(e,x) := (I_eF^{\rm P}_{\mu\nu})(x)e^\nu = A^{\rm P}_\mu (x) 
+ \partial_\mu (I_eA^{\rm P}_\nu)(x)e^\nu 
\equiv (J_{e})_\mu{}^\nu A^{\rm P}_\nu 
\eea 
with $I_e$ and $J_e$ defined in \eref{I}, \eref{Jx}. Its \tpf\ arises
by integration over the Proca \tpf, that results in a multiplication
of the Proca \tpk\ \eref{PP} with the matrices $J_e(p)$ defined in \eref{Je}:
\bea\label{E=JJ}
\merw m{A_\mu(-e),A_{\mu'}(e')}(p)=-E(e,e')(p)_{\mu\mu'}\equiv
- J_e(p)_\mu{}^{\nu}J_{e'}(p)_{\mu'}{}^{\nu'}\eta_{\nu\nu'}. \quad
\eea
Explicitly, 
\bea
\label{E}
E(e,e')(p)_{\mu\mu'}  =\Big[\eta_{\mu\mu'} 
- \frac{e'_\mu p_{\mu'}}{(pe')_+} - \frac{p_\mu e_{\mu'}}{(pe)_+} 
+ \frac{(ee')p_\mu p_{\mu'}}{(pe)_+(pe')_+}\Big].
\eea
We have chosen the string $-e$ in the left argument in \eref{E=JJ} in
order to avoid denominators $1/(-pe)_+ = -1/(pe)_-$ if the argument were
$+e$. In particular, $E(e,e)(p)$ is the momentum space version of the
operator $E_e$ in \eref{coup}.

It follows from \eref{E=JJ}
\bea
\label{aa} 
\merw m{(\partial A)(-e),(\partial A)(e')}(p)=-p^\mu p^\nu
E(e,e')(p)_{\mu\nu} = m^2 \Big(1-m^2\frac{(ee')}{(ep)_+(e'p)_+}\Big).
\eea
Therefore, the field $a(e,x):=-m\inv \partial_\mu A^\mu(e,x)$ is
regular at $m=0$ and converges to the canonical scalar field $\varphi$.  

For the \sloc\ Maxwell potential \eref{Imax}, which has no covariant
\ploc\ potential on the Hilbert space, one has to compute the \tpf\ by
integration over the field strength whose \tpk\ is 
$\merw{0}{F_{\mu\nu}, F_{\kap\lambda}} = 
-p_\mu p_\kap\eta_{\nu\lambda}+ p_\nu p_\kap\eta_{\mu\lambda}
+p_\mu p_\lambda\eta_{\nu\kap} -p_\nu p_\lambda\eta_{\mu\kap}$. This gives 
the same formula \eref{E=JJ} except that the mass is zero and $p^2=0$.  

This continuity property does not persist at $s>1$, see
\cite{MRS1,MRS2}, where the decoupling of the lower 
helicities is more subtle than at $s=1$. 

\subsection{Two-point functions of infinite-spin fields}
\label{s:tpf-inf}

The \tpk s of infinite spin fields \eref{infspin} given in terms of their
intertwiners are 
\bea
\label{tpf-inf}
\merw{0}{\phi_u(-e),\phi_u(e')}(p) = \int
d\mu_\kappa(\vec k) \, \overline{u(-e,p)(\vec k)} u(e',p)(\vec k).
\eea
For the standard field $\phi^{\kappa(0)}=\Phi^{\kappa(0)}$, we get 
\begin{propo}\label{p:tpf-std}  (see also \cite[Chap.~4.1]{K})
The \tpk\ of $\phi^{\kappa(0)}(e,x)$ is
\bea
\label{tpf-std}
\merw{0}{\phi^{\kappa(0)}(-e),\phi^{\kappa(0)}(e')}  
= \ol{\omega^\kappa(-e,p)}\omega^\kappa(e',p)\cdot
J_0\big(\kappa R_{e,e',p}\big),
\eea
where $q_e(p):=\frac e{(ep)_+}$ and $R_{e,e',p}^2=-(q_e(p)-q_{e'}(p))^2$.
\end{propo}
Notice that $q_e-q_{e'}$ is orthogonal to $p$ because $(pq_e)=(pq_{e'})=1$, 
hence $q_e-q_{e'}$ is spacelike and the argument of the Bessel
function is real. The distributional nature of $R_{e,e',p}$ is absorbed by
the K\"ohler factors, see \pref{p:k-bd}.
 
Proof: Recall the definition of the Bessel functions 
\bea
\label{bessnu}
\frac1{2\pi}\int d\varphi \, e^{i\nu\varphi}
e^{iz\cos\varphi} = i^\nu J_\nu (z) = i^{-\nu} J_{-\nu}(z).
\eea
Thus, with $\wt u^{\kappa}$ given by \eref{ut0}, the integral in
\eref{tpf-inf} can be performed:
\bea\label{int-bess}\int d\mu_\kappa(\vec k) \,e^{i\frac{(eE_p(\vec
    k))}{(ep)_+}} e^{-i\frac{(e'E_p(\vec
    k))}{(e'p)_+}}  = \int d\mu_\kappa(\vec k) \,e^{-i\vec
  k\cdot \vec f} = J_0(\kappa\vert \vec f\vert)
\eea
where $\vec f$ is the transverse ($1$-$2$-)part of the four-vector 
$$f= B_p\inv\Big(\frac{e}{(ep)_+} - \frac{e'}{(e'p)_+}\Big) 
=B_p\inv(q_e-q_{e'}).$$
Because $(fp_0)=0$, one has $f^2=(fp_0)(f\xi_0) -\vert \vec f\vert^2 =
-\vert \vec f\vert^2$. The claim follows by multiplying with the
K\"ohler factors. \qed 
\begin{propo}
\label{p:Phirr'} 
The \tpf s among $\Phi^{\kappa(r)}$ are given by their \tpk s (no
respective sums if $r=0$ or $r'=0$)
\bea\notag\merw{0}{\Phi^{\kappa(r)}(-e,v),\Phi^{\kappa(r')}(e',v')} = 
\ol{\omega^\kappa(-e,p)}\omega^\kappa(e',p)\cdot\sum\nolimits_{\nu=\pm
r}(v\Eps_\pm)^r\sum\nolimits_{\nu'=\pm r'}
(v\Eps'_\pm)^{r'}\cdot\\\notag
\hspace{30mm}\cdot e^{-i(\nu+\nu')\alpha_{e,e',p}}\cdot
i^{\nu+\nu'}J_{\nu+\nu'}\big(\kappa R_{e,e',p}\big).
\eea
Here, $R_{e,e',p}\bpm\cos\alpha_{e,e',p}\\[-.5mm]\sin\alpha_{e,e',p}\epm$
parametrizes the $1$-$2$-part of $B_p\inv (q_e(p)-q_{e'}(p))$.
\end{propo}
Proof: By a direct computation, using \eref{UrEps} and \eref{bessnu}. 
\qed
\begin{remark}\label{kappa0}
The fields $\Phi^{\kappa(r)}$ mutually decouple in the
limit $\kappa\to0$, because $J_\nu(0) = \delta_{\nu,0}$. Their limits
as $\kappa\to0$ coincide with the massless potentials $A^{(r,r)}$,
i.e., the infinite-spin field ``decays'' into a direct sum of massless
fields of every helicity \cite[Chap.~4]{K}.  
\end{remark}

\subsection{Massive fields of finite spin: definitions and properties}
\label{s:massive}
Our aim is to approximate \eref{tpf-std} by higher-spin
generalizations of \eref{E=JJ} and \eref{aa} in the \pll. We take
stock of the relevant results in \cite{MRS1,MRS2}, and supplement it
by the crucial recursive formula \pref{p:asc}.

We are going to work with the string-localized fields\footnote{The
  first superscript $s$ was suppressed in \cite{MRS1,MRS2}, where we
  worked at fixed $s$.}
$a^{(s,r)}_{\mu_1\dots\mu_r}(e,x)$ defined in \eref{ass} and \eref{desc}, 
and the \sloc\ fields $A^{(s,r)}(e,x)$ defined in \eref{Asr}.
All these fields are derived from the total field strengths
$F^{{\rm P}(s)}_{[\mu_1\nu_1]\dots[\mu_s\nu_s]}(x)$ of the \ploc\ massive 
``Proca'' fields $A^{{\rm P}(s)}_{\mu_1\dots\mu_s}(x)$ of spin $s$ \cite{Fz}. 
We start with the definition
\bea
\label{ass}
a^{(s,s)}_{\mu_1\dots\mu_s}(e,x):= \big( (I_e)^s
F^{{\rm P}(s)}_{[\mu_1\nu_1]\dots[\mu_s\nu_s]}\big)(x) \cdot 
e^{\nu_1}\dots e^{\nu_s}
\eea
and the descending recursion
\bea\label{desc}
-m\cdot a^{(s,r)}_{\mu_1\dots\mu_r}(e,x):= \partial^\mu
a^{(s,r+1)}_{\mu\mu_1\dots\mu_r}(e,x).
\eea
$a^{(s,s)}$ differs from $A^{{\rm P}(s)}$ by derivatives of $a^{(s,r)}$ 
($r<s$) \cite[Prop.~3.5]{MRS1}, so that it is also a potential for the
field strength $F^{{\rm P}(s)}$. Unlike the Proca potential, the field
strength and hence also the
string-localized potential $a^{(s,s)}$ and its escort fields
$a^{(s,r)}$ are regular in the limit $m\to0$ at fixed spin $s$. The
discrepancy between these limits and potentials $A^{(r)}$ for the
massless field strengths $F^{(r)}$ was studied in \cite{MRS1}.

Let us identify some general patterns. The composition of the operations
$$\hbox{(contraction with $e$)}\circ\hbox{(string integration)}\circ\hbox{(curl)},$$
applied to each index of $A^{{\rm P}(s)}$ in the definition
\eref{ass}, is the matrix operator $(J_e)_\mu{}^\nu = \delta_\mu^\nu +
I_ee^\nu\partial_\mu$ already seen in \eref{Jx} and
\eref{Iproca}. Thus, \eref{ass} and \eref{desc} become
\bea
\label{aA} \notag
a^{(s,s)}_{\mu_1\dots\mu_s}(e,x)&=&
(J_e)_{\mu_1}{}^{\nu_1}\dots(J_e)_{\mu_s}{}^{\nu_s} 
A^{{\rm P}(s)}_{\nu_1\dots\nu_s}(x),\\[1mm] 
a^{(s,r)}_{\mu_1\dots\mu_r}(e,x)&=&
(J_e)_{\mu_1}{}^{\nu_1}\dots(J_e)_{\mu_r}{}^{\nu_r} 
(me^{\nu_{r+1}}I_e)\dots(me^{\nu_{s}}I_e)A^{{\rm P}(s)}_{\nu_1\dots\nu_s}(x),
\eea
Because $J_{\lambda e} = J_e$ ($\lambda>0$), all $a^{(s,r)}(e,x)$ are
homogeneous distributions in $e$. 

The following formula for the $e$-dependence of $a^{(s,s)}(e,x)$ was
displayed in \cite[Cor.~3.3]{MRS1} with an erroneous factor $m\inv$: 
\bea
\label{deas} 
\partial_{e^\mu} a^{(s,s)}_{\mu_1\dots\mu_s}(e,x) = \sum\nolimits_{i=1}^s
\partial_{\mu_i} I_e a^{(s,s)}_{\dots\mu_{i-1}\mu\mu_{i+1}\dots}(e,x).
\eea
It follows from the definition of $a^{(s,s)}$ and the identities
\eref{deJ} and \eref{Iinv}.
\begin{propo}\label{p:asc} The descending defining recursion
  \eref{desc} is inverted by the ascending recursion, involving the
  variation of the direction $e$, 
\bea\label{asc}
-m(s-r) \cdot a^{(s,r+1)}_{\mu\mu_1\dots\mu_r}(e,x) =
\sum\nolimits_{i=1}^r \partial_{\mu_i}
a^{(s,r)}_{\dots\mu_{i-1}\mu\mu_{i+1}\dots} + (e\partial_x)
\partial_{e^\mu} a^{(s,r)}_{\mu_1\dots\mu_r}(e,x).
\eea
In particular, the r.h.s.\ is completely symmetric in
$\mu,\mu_1,\dots,\mu_r$. 
\end{propo}
This result conversely exhibits also the $e$-dependence of 
$a^{(s,r)}_{\mu_1\dots\mu_r}(e,x)$ via \eref{dear}.

Proof (\cite{G17}):
Apply $\partial^{\mu_s}\dots\partial^{\mu_{r+1}}$ to \eref{deas} and
use the defining recursion \eref{desc}. This gives 
\bea
\label{dear}
\partial_{e^\mu} a^{(s,r)}_{\mu_1\dots\mu_r}(e,x) = m(s-r) \cdot I_e
a^{(s,r+1)}_{\mu\mu_1\dots\mu_r}(e,x) + \sum\nolimits_{i=1}^r \partial_{\mu_i}
I_e a^{(s,r)}_{\dots\mu_{i-1}\mu\mu_{i+1}\dots},
\eea
which can be solved for $a^{(s,r+1)}_{\mu\mu_1\dots\mu_r}(e,x)$ using \eref{Iinv}. 
\qed

We also report the identities
\begin{propo}
\label{p:trax} (see \cite[Prop.~3.6]{MRS1}) 
For $r\geq 2$, resp.\ $r\geq0$ 
\bea
\label{trax}
\eta^{\kap\lam} a^{(s,r)}_{\kap\lam\mu_3\dots\mu_r} = 
-a^{(s,r-2)}_{\mu_3\dots\mu_r}, 
\qquad e^\kap a^{(s,r)}_{\kap\mu_2\dots\mu_r}=0.
\eea
\end{propo}
Proof: By inspection of \eref{aA}, using that $A^{{\rm P}(s)}$ is
traceless and conserved. \qed 

We now turn to \tpf s.

The intertwiner for the $(m,s)$ Proca potential is
given by the $s$-fold tensor product of the standard intertwiner for
spin $1$ \cite{W}, preceded by the projection onto the traceless
symmetric subrepresentation (= spin $s$ representation) of the tensor
product of spin 1 representations of the little group $SO(3)$: 
\bea
u_n^{{\rm P}(s)}(p) = (B_pE_3)^{\otimes s} T_n,
\eea
where $T_n$, $n=1,\dots,2s+1$, is an
orthonormal basis of traceless symmetric tensors in
$(\RR^3)^{\otimes s}$, and $E_3:\RR^3\to\RR^4$ the standard embedding into
Minkowski space. 

The resulting \tpk\ of the Proca field is (in the notations introduced
in \sref{s:tensor1} and \sref{s:basics})
\begin{propo}\label{p:AA} (see \cite{G} and \cite[Sect.~2.1]{MRS1})
\bea\label{AA}
\merw{m}{A^{{\rm P}(s)}(v),A^{{\rm P}(s)}(v')}= (-1)^s 
\sum\nolimits_{2n\leq  s}\beta_n^s \cdot
[(v\pi v)(v'\pi v')]^n\cdot(v\pi v')^{s-2n}.
\eea
Here, $\pi_{\mu\nu}=\eta_{\mu\nu}-\frac{p_\mu p_\nu}{m^2}$, and $(v\pi
v)$ etc.\ are the contractions with $v$ resp.\ $v'$.
The coefficients
$\beta^s_n=\frac{1}{4^nn!}\frac{s!}{(s-2n)!}\frac{1}{(\frac12-s)_n}$
(ensuring the tracelessness) are the coefficients of the
hypergeometric function 
\bea\label{hgf}
F_s(z)\equiv \sum\nolimits_{2n\leq  s}\beta_n^s z^n =
{}_2F_1\Big(\frac {-s}2,\frac {1-s}2;\frac 12-s;z\Big).
\eea
This function is in fact a polynomial of order $\lfloor\frac
s2\rfloor$, because either $\frac{-s}2$ or $\frac{1-s}2$ is a non-positive
integer.
\end{propo}
The \tpk\ of the massive spin $s$ string-localized
potential $a^{(s,s)}$ follows from the definition \eref{ass}:
\newpage
\begin{propo}\label{p:tpf-ss} (see \cite[Sect.~3]{MRS1})
\bea\label{tpf-ss}
\merw{m}{a^{(s,s)}(-e,v),a^{(s,s)}(e',v')}= (-1)^s 
\sum\nolimits_{2n\leq  s}\beta_n^s \cdot
[(vEv)(v'E''v')]^n\cdot(vE'v')^{s-2n}.\eea
Here, $E\equiv E(e,e)(p)$, $E'\equiv E(e,e')(p)$, $E''\equiv E(e',e')(p)$ as
in \eref{E=JJ}.
\end{propo}
From \eref{tpf-ss}, one gets the correlations of all escort fields 
$a^{(s,r)}$ by descending in $r$ with the defining recursion
\eref{desc}. It turns out to be more convenient to descend
directly to $r=0$:
$a^{(s,0)}(e,x)=\frac1{s!}(-m)^{-s}(\partial_x\partial_v)a^{(s,s)}(e,x,v)$,
and then use the ascending recursion \eref{asc}. This strategy will allow to
study limits of $a^{(s,r)}_{\mu_1\dots\mu_r}$ for fixed $r$ while $s$ increases.

Taking the divergence in all indices via
$(\partial_x\partial_v)=-i(p\partial_v)$ and
$(\partial_{x'}\partial_{v'})=+i(p\partial_{v'})$, just amounts to
putting $v=v'=p/m$ in the \tpk\ \eref{tpf-ss}. One gets 
the \tpk\ of the string-localized ``scalar'' escort field 
$a^{(s,0)}$:
\begin{propo}\label{p:tpf-00} (see \cite[Sect.~3]{MRS1})
\bea\label{tpf-00}
\merw{m}{a^{(s,0)(-e)},a^{(s,0)}(e')} =
(-1)^sm^{-2s}\sum\nolimits
\beta^s_n\cdot[(pEp)(pE''p)]^n\cdot(pE'p)^{s-2n},
\eea
where $(pEp)=E(e,e)(p)=
-m^2(1-m^2q_e^2)$, and similar for $(pE'p)$, $(pE''p)$. 
\end{propo}
We rewrite \eref{tpf-00} as 
\bea\label{tpf-hgf}
\merw{m}{a^{(s,0)}(-e),a^{(s,0)}(e')} = (1-m^2(q_{e}q_{e'}))^s \cdot
F_s(z_m(q_{e},q_{e'})) \equiv P^s_m(q_{e},q_{e'})
\eea
where $z_m(q,q')=\frac{(1-m^2q^2)(1-m^2q'^2)}{(1-m^2(qq'))^2}$. Notice
that $P^s_m(q_e,q_{e'})$ is a polynomial in $q_e,q_{e'}$, hence immediately
well-defined as a distribution. 

From this, one may obtain the \tpk s for $a^{(s,r)}$ by ascending
with \eref{asc}, using \eref{Je} and \eref{deJ} in momentum space. 

We have also defined \sloc\ fields $A^{(s,r)}$ that decouple in the
massless limit at fixed $s$, and become potentials for the massless
field strengths $F^{(r)}$ of helicity $\pm r$: 
\begin{defn}\label{d:Asr} (see \cite[Prop.~3.8]{MRS1})
\bea\label{Asr}
A^{(s,r)}(e,x,v):= N^{(s,r)}_0
\cdot \sum\nolimits_{2k\leq r}\gamma^r_k \cdot
(-E_e(v))^k \, a^{(s,r-2k)}(e,x,v)
\eea
with $E_e$ as in \eref{coup}. The coefficients are
$\gamma^r_k=\frac{1}{4^kk!}\frac{r!}{(r-2k)!}\frac{1}{(1-r)_k}$, and 
$N^{(s,r)}_0 := \big[\binom sr \frac{\Gamma(\frac12+s)\Gamma(1+r)}
{\Gamma(\frac12+\frac{r+s}2)\Gamma(1+\frac{r+s}2)}\big]^{\frac12}$.
\end{defn}
\begin{propo}\label{p:Asr} (see \cite[Cor.~3.10]{MRS1}) In the limit
  $m\to0$ at fixed 
  $s$ and $r$, the \sloc\ fields $A^{(s,r)}(e,x)$ mutually decouple, and are
  potentials for the \ploc\ massless field strengths $F^{(r)}(x)$ associated
  with the Wigner representations of helicities $\pm r$. In
  particular, at $m=0$ they no longer depend on $s$, they are
  traceless, conserved, and satisfy $e^\mu A^{(s,r)}_{\mu\mu_2\dots\mu_r}=0$.
\end{propo}

\section{The \pll}
\label{s:PLL}
\setcounter{equation}{0}

After these preparations, we turn to the \pll. The limit of the
``scalar'' escort fields $a^{(s,0)}(e,x)$ is rather easy. It relies on
two lemmas.  
\begin{lemma}\label{l:erd} For $s\in \NN$, the identity of polynomials
  of degree $\lfloor\frac s2\rfloor$
$$
{}_2F_1\Big(\frac {-s}2,\frac {1-s}2;\frac 12-s;z\Big) = 
\frac{\Gamma(1+\frac s2)\Gamma(\frac12+\frac s2)}{\Gamma(\frac12+s)}
\cdot {}_2F_1\Big(\frac {-s}2,\frac {1-s}2;1;1-z\Big)
$$
holds. We henceforth abbreviate this identity as
\bea\label{FG}
F_s(z)=F_s(1)\cdot G_s(1-z),
\eea 
in accord with the previous notation \eref{hgf}. In particular, 
$F_s(1)=\frac{\Gamma(1+\frac s2)\Gamma(\frac12+\frac s2)}
{\Gamma(\frac12+s)}$ $\equiv(N_0^{(s,0)})^{-2}$. For large $s$,
this decays asymptotically like $F_s(1)\approx 2^{-s}\sqrt{\pi s}$. 
\end{lemma}
Proof: The identity is the formula 2.9(43) in \cite{Erd} with parameters
$a=\frac{-s}2$, $b=\frac{1-s}2$, $c=\frac12-s$. The value $F_s(1)$
follows because $G_s(0)=1$. The asymptotic form follows by Stirling's
approximation of the $\Gamma$ function. 
\qed
\begin{lemma}\label{l:bess} 
In the limit $s\to\infty$, the pointwise limit holds
\bea\label{bess}
\lim_{s\to\infty} G_s\Big(-\frac{u^2}{s^2}\Big) \equiv
\lim_{s\to\infty} 
{}_2F_1\Big(\frac {-s}2,\frac {1-s}2;1;-\frac{u^2}{s^2}\Big) = J_0(u).
\eea
\end{lemma}
Proof: The power series expansion reads
$$G_s\Big(-\frac{u^2}{s^2}\Big) = 
\sum\nolimits_{k\geq0} \frac{(-s)_{2k}}{4^kk!^2}\Big(-\frac{u^2}{s^2}\Big)^k
\to
\sum\nolimits_{k\geq0}\frac{1}{k!^2}\Big(-\frac{u^2}{4}\Big)^k=J_0(u),$$
where the limit of the coefficients is taken separately for each $k$. 
The pointwise convergence in $u$ follows by absolute convergence of the
sums. \qed

Now we turn to the fields. 

\begin{propo}\label{p:a-lim} In the \pll, the \tpk\ of
  the rescaled scalar escort fields $N_0^{(s,0)} \cdot a^{(s,0)}(e)$ 
  converges to $J_0(\kappa \sqrt{-(q_{e}(p)-q_{e'}(p))^2})$. Since
  $J_0(z)$ is a power 
  series in $z^2$, the \tpk\ is a power series in $q_{e}$ and $q_{e'}$. 
  The convergence is pointwise, i.e., it holds formally for fixed values
  of $q_{e}$ and $q_{e'}$, and more precisely for fixed test functions
  in $e$, $e'$ and $p$, on whose support $(q_{e}(p)-q_{e'}(p))^2$ is bounded.  
\end{propo}
Proof: Using \lref{l:erd}, we rewrite \eref{tpf-hgf} as 
\bea\label{aa=FG}
\merw{m}{a^{(s,0)}(-e),a^{(s,0)}(e')} = (1-m^2(q_{e}q_{e'}))^s \cdot F_s(1)
G_s(1-z_m(q_{e},q_{e'})).
\eea
The prefactor $(1-\frac{\kappa^2(qq')}{s(s+1)})^s$ converges
separately to $1$ by Euler's formula. The claim then follows by \lref{l:bess}.
\qed

\newpage

Comparing the limit obtained in \pref{p:a-lim} with the  
\tpk\ \eref{tpf-std} of the standard \sloc\ field $\phi^{\kappa(0)}$,
one notes that the K\"ohler factors are missing. But they can be
produced by applying the operators $(1-m\sqrt{-e^2}I_e)^s$ before the
limit is taken. Namely, by Euler's formula, 
$\omega^\kappa(e,p)= e^{-i\kappa\frac{\sqrt{-e^2}}{(ep)_+}}$
is the limit of $\big(1-im\frac{\sqrt{-e^2}}{(ep)_+}\big)^s$, and 
$\frac{i}{(ep)_+}$ is the momentum space version of the
string-integration $I_e$.
\begin{coro}\label{c:u0}
The standard \sloc\ field $\phi^{\kappa(0)}(e,x)$ is, up to unitary
equivalence, the \pll\ of  
$N_0^{(s,0)} \cdot (1-m\sqrt{-e^2}I_e)^sa^{(s,0)}(e,x)$.
\end{coro}
\begin{remark}\label{r:conv}
The convergence of the \tpk s is much easier to see than that of the
intertwiners, because the former is basis independent. The
reason is that ``convergence'' of vectors on different Hilbert spaces
makes only sense with a suitable inductive limit (a sequence of 
embeddings of the Hilbert spaces). For the case at hand, this 
inductive limit of the representation spaces of the massive little
group $SO(3)$ is described in \aref{a:ind}, in such a way that the
matrix elements converge to a representation of the massless little
group $E(2)$, and this extends to the induced Wigner representation \cite{McK}. 

With the given inductive identification of bases, one should be able to 
prove the convergence of intertwiners up to unitary equivalence. We refrain
from doing this because the \tpf\ uniquely specifies a free field, and
hence we may conclude the convergence of the intertwiners and
of the fields up to unitary equivalence. In this sense, we may say
\bea\label{a-lim}
\lim_\kappa N_0^{(s,0)}\cdot (1-m\sqrt{-e^2}I_e)^sa^{(s,0)}(e,x) =
\phi^{\kappa(0)}(e,x). 
\eea
\end{remark}
\begin{propo}\label{p:ar-lim} The fields $N_0^{(s,0)}\cdot
  (1-m\sqrt{-e^2}I_e)^sa^{(s,r)}(e,x)$ converge in the \pll\ (in the
  same sense as specified in \rref{r:conv}) to the infinite-spin
  fields $\phi^{\kappa(r)}(e,x)$ defined in \dref{d:ur-def}.
\end{propo}
Proof: (i) The recursion \eref{asc} for $a^{(s,r)}$ implies the 
recursion 
$$
-\kappa \wt\phi^{\kappa(r+1)}(e,x,v) =
\big(r\cdot(v\partial_x)+(e\partial_x)(v\partial_e)\big)
\wt\phi^{\kappa(r)}(e,x,v)
$$
for the limits of $N_s^{(s,0)}\cdot a^{(s,r)}$, because for each fixed
$r$, $m(s-r)$ can be replaced by $\kappa$. Restoring the K\"ohler
factors of the intertwiners by means of the operators
$(1-m\sqrt{-e^2}I_e)^s$ as in \pref{p:a-lim}, implies the recursion 
\eref{phi-rec} for the limits of $N_0^{(s,0)}\cdot
  (1-m\sqrt{-e^2}I_e)^sa^{(s,r)}(e,x)$. The claim follows,
  because \eref{phi-rec} defines $\phi^{\kappa(r)}$. 
\qed

We now study the \pll\ of the fields $A^{(s,r)}$, defined in \dref{d:Asr}. 
Using the trace identity in \eref{trax}, we rewrite \eref{Asr} as 
\bea\label{Asr-cp}
A^{(s,r)}(e,x) = \frac{N^{(s,r)}_0}{N^{(s,0)}_0}\cdot
\Big( S_r\sum\nolimits_{2k\leq r}\gamma^r_k\cdot 
((J_e\otimes J_e)\eta\circ\eta^t)^{\otimes k} \Big) 
\cdot N_0^{(s,0)}\, a^{(s,r)}(e,x), \quad
\eea
regarding $a^{(s,r)}$ as a tensor in $(\RR^4)^{\otimes r}$ of which
pairs of indices are contracted with $\eta$ and restored by 
$E_e=(J_e\otimes J_e)\eta$, and the result in $(\RR^4)^{\otimes r}$ is
symmetrized.  
\begin{coro}\label{c:Ar-lim}
In the \pll, the fields $(1-m\sqrt{-e^2}I_e)^sA^{(s,r)}(e,x)$ converge 
(in the same sense as specified in \rref{r:conv}) to the infinite-spin
fields $\Phi^{\kappa(r)}(e,x)$ defined in \dref{d:Ur-def}.
\end{coro}
Proof: One easily sees from the formula given in \dref{d:Asr} that 
the prefactor $\frac{N^{(s,r)}_0}{N^{(s,0)}_0}$
converges to $2^{r/2}$ as $s\to\infty$. The operator does not depend on
$s$ and $m$ and commutes with $(1-m\sqrt{-e^2}I_e)^s$, and 
$N_0^{(s,0)}\cdot (1-m\sqrt{-e^2}I_e)^sa^{(s,r)}(e,x)$
converges to $\phi^{\kappa(r)}$ by \pref{p:ar-lim}. Thus, the limit of
\eref{Asr-cp} is equivalent to \eref{Ur-def}. \qed

\section{Stress-energy tensors}
\label{s:sets}
\setcounter{equation}{0}
In \cite{MRS1}, we have introduced several \set s for the $(m,s)$
fields that all yield the correct infinitesimal generators 
\bea\label{PM}
P_\sig = \int_{x_0=t} d^3\vec x \,
T_{0\sig}, 
\qquad
M_{\sig\tau}  = \int_{x_0=t} d^3\vec x \, (x_\sig T_{0\tau} - x_\tau
T_{0\sig})
\eea
of the Poincar\'e group. They differ by derivative terms that
vanish upon the integrations \eref{PM}. We display here: the
point-localized ``reduced'' \set\
\bea\label{Tred}
T^{(s)\rm red}_{\rho\sig}(x)= - \frac 14(-1)^s 
\wick{A^{{\rm P}(s)\mu_1\dots\mu_s}\lrd\rho\lrd\sig
A^{{\rm P}(s)}_{\mu_1\dots\mu_s}}(x) 
+ \partial^\mu \Delta T^{(s)\rm red}_{\rho\sig;\mu}
\eea
and the string-localized ``regular'' \set\
\bea\label{Treg}
t^{(s)\rm reg}_{\rho\sig}(e_1,e_2,x) &=& \sum\nolimits_{r\leq s}\binom sr
t^{(s,r)}_{\rho\sig}(e_1,e_2,x), \\ \notag
t^{(s,r)}_{\rho\sig}(e_1,e_2,x) &=& - \frac 14(-1)^r 
\wick{a^{(s,r)\mu_1\dots\mu_r}(e_1)\lrd\rho\lrd\sig
a^{(s,r)}_{\mu_1\dots\mu_r}(e_2)}(x) 
+ \partial^\mu \Delta t^{(r)\rm reg}_{\rho\sig;\mu}.
\eea
The derivative terms $\partial^\mu \Delta T_{\rho\sig;\mu}$ do not
affect the momentum generators, but they have to be added to get the
correct infinitesimal Lorentz transformations. Explicit
expressions can be found in \cite{MRS1}. We do not need them at
this point, see however \xref{x:x}. The tensors $t^{(s,r)}$ are
separately conserved, but only 
their sum \eref{Treg} generates the correct Poincar\'e transformations.  

We have also given massless stress-energy tensors
\bea\label{Tml}
T^{(s),m=0}_{\rho\sig}(e_1,e_2,x) \!\!&=&\!\! \sum\nolimits_{r\leq s}
\binom sr T^{(s,r),m=0}_{\rho\sig}(e_1,e_2,x), \\ \notag
T^{(s,r),m=0}_{\rho\sig}(e_1,e_2,x) \!\!&=&\!\! - \frac 14(-1)^r 
\wick{A^{(s,r)\mu_1\dots\mu_r}(e_1)\lrd\rho\lrd\sig
A^{(s,r)}_{\mu_1\dots\mu_r}(e_2)}(x) + \partial^\mu \Delta
T^{(s,r),m=0}_{\rho\sig;\mu},
\eea
that, in spite of the definition of the \sloc\ fields $A^{(s,r)}$ as
massless limits of fields on the massive Fock space of spin $s$ (in
terms of $a^{(s,r'\leq r)}$, see \dref{d:Asr}), do not depend on $s$. 
At $m=0$, \eref{Tml} describes the decoupling of the massive spin $s$
representation into a direct sum of helicity representations \cite{MRS1,MRS2}. 

Let us discuss the possible role of these tensors in the \pll. The
\ploc\ reduced \set\ $T^{(s)\rm red}$ does not admit a massless
limit because of inverse powers $m^{-4s}$ in the \tpf. 
The \sloc\ regular \set\ $t^{(s)\rm reg}$ becomes in the \pll\ an
infinite sum over $r\leq s\to\infty$ of terms $t^{(s,r)}$, that each
converge to zero (due to the explicit factor $F_s(1)\approx 2^{-s}$
present in every \tpf\ $(\Omega,a^{(s,r)}a^{(s,r')}\Omega)$). Yet, the
sum is not zero and is still a valid \set, but it cannot be expressed
as a sum of limits of $t^{(s,r)}$. (These and other interesting 
features are nicely illustrated by the expectation values of the
energy density and the pressure in thermal states at inverse
temperature $\beta$, cf.\ \sref{s:KMS}. E.g., while the contribution
of each $r$ to the thermal energy goes to zero, the sum over $r$
diverges like $2s+1$.)     

\eref{Tml} seems to be better suited for the \pll, because each term
$T^{(s,r)}$ has a limit. At fixed $s$, the massless \set\ $T^{(s),m=0}$ 
is the limit of massive conserved tensors $T^{(s),m}$ \cite{MRS1}. 
The latter differ from $t^{(s)\rm reg}$, apart from irrelevant terms
that do not affect the Poincar\'e generators, by 
further terms that do disturb the generators, and that decay like
$O(m)$ at fixed $s$ \cite[Props.~4.5 and 4.6]{MRS1}. But such terms
may grow with $s$, so that it is difficult to keep control, whether
the \pll\ produces the correct generators. 

Our main result in this section computes the \set\ directly in the
infinite-spin representation: 
\begin{propo} \label{p:set}
Let $\Phi^{\kappa(r)}(e,x)$ be the \sloc\ fields defined in
\dref{d:Ur-def} (that may be obtained as \pll s by \cref{c:Ar-lim}). 
The conserved symmetric tensor 
\bea\label{Tinf}
T^{\kappa}_{\rho\sig}(e,x) &=& \sum\nolimits_{r=0}^\infty
T^{\kappa (r)}_{\rho\sig}(e,x), \\ \notag
T^{\kappa (r)}_{\rho\sig}(e,x) &=& - \frac 14(-1)^r \wick{
\Phi^{\kappa(r)\mu_1\dots\mu_r}(e) \lrd\rho\lrd\sig
\Phi^{\kappa(r)}_{\mu_1\dots\mu_r}(e)}(x) 
+ \partial^\mu \Delta T^{\kappa(r)}_{\rho\sig;\mu}(e,x)
\eea
is a \sloc\ \set\ for the infinite-spin representation. An expression for
$\Delta T^{\kappa(r)}_{\rho\sig;\mu}$ will be given in \eref{DelTinf}. 
\end{propo}
The proof in \sref{s:proof} exhibits $\Delta T^{\kappa(r)}$ as a sum of two
pieces. The ``second piece'' $\Delta_2T^{\kappa(r)}$ is absent at
finite $s$ and must be identified with the accumulation of the
above-mentioned uncontrolled errors.  
\begin{remark}\label{r:warning} In order to get the correct
  generators, the two \sloc\ fields in the Wick product have to be
  taken with $e_1=-e_2$, see \rref{r:ee}. Mund has recently shown that
  the Wick product with parallel strings is well-defined
  as a distribution in $x$ and $e$.
  
  The problematic issue with \eref{Tinf} is instead the infinite sum
  over $r$. Because the fields $\Phi^{\kappa(r)}$ do not decouple
  (see \eref{coup}), correlation functions and matrix elements
  involving $T^\kappa$ may be divergent sums.  
  Recall from \rref{r:sharp} that each $T^{\kappa(r)}$ will have a
  non-vanishing expectation value in a state with sharp magnetic
  quantum number $n\in\ZZ$. The consequences of this feature for \tpf
  s and commutators of $T^\kappa$ with the fields $\Phi^{\kappa(r)}$
  will be sketched in \sref{s:props}. 
\end{remark}

\subsection{Quantum \set s}
\label{s:qsets}

We obtained \pref{p:set} with a new systematic strategy to find \set s
for free quantum fields, that does not refer to a classical action principle. 
Instead, it is intrinsically based on the Wigner representation
theory, along with a choice of intertwiners that allow to ``decompose''
the (global) generators into integrals over localized densities. 

We first outline the general strategy, that is flexible enough
to include also point- and \sloc\ finite-spin tensor fields and Dirac
fields. The \set s \eref{Tred}, \eref{Treg} and \eref{Tml} could have
been found by this strategy.

Applied to the infinite-spin case (\sref{s:proof}), it does not use
the Pauli-Lubanski approximation, i.e., it proceeds directly in the
limit, using just the results of \pref{p:Ur}. 

To keep the argument transparent, we present only the case of bosonic
hermitean fields, where
the $u$- and $v$-intertwiners multiplying creation and annihilation
operators are complex conjugates of each other. 

We start with the familiar global ``second quantization'' formula for
the momentum operator  
\bea\label{P}
P_\sig = \int d\mu_m(p)\, \sum\nolimits_n p_\sig a^*_n(p) \,a_n(p),
\eea
where the sum extends over an orthonormal basis of the representation
space $\HH_d$ of the unitary representation
$d$ of the little group. We write this as 
$$P_\sig = \int d\mu_m(p_1)d\mu_m(p_2)\, 
\sum\nolimits_{n_1n_2}p_{1\sig}a^*_{n_1}(p_1)\,
\delta_{n_1n_2}\,(2\pi)^3\delta(\vec p_1-\vec p_2)(p_{10}+p_{20}) \,a_{n_2}(p_2),$$
and insert $(2\pi)^3\delta(\vec p_1-\vec p_2)=\int
d^3\vec x\,e^{-i(\vec p_1-\vec p_2)\vec x}=\int d^3\vec x\,e^{i(p_1-p_2)x}$, 
which is independent of $x^0$. Separating the factors that depend on
$p_1$ and on $p_2$, respectively, and interchanging the $x$- and
$p$-integrations, one gets an $\vec x$-integral over the product of
(derivatives of) two expressions 
$\int d\mu_m(p_1)\,e^{ip_1x}a_{n_1}^*(p_1)$ and
$\int d\mu_m(p_2)\,e^{-ip_2x}a_{n_2}(p_2)$. These are of course {\em not} the
creation and annihilation parts of a local and covariant quantum field,
by the well-known problem of the nonlocal Wigner ``rotations'', that is
the reason why one has to use intertwiners in Wigner quantization \cite{W}. 

So, let there be a (possibly reducible) representation $D$ of the
Lorentz group and intertwiners $u_{Mn}(p)$ satisfying
\bea\label{itw-gen}
u(\Lambda p) = \big(D(\Lambda)\otimes d(W_{\Lambda,p})\big) u(p)
\qquad (\Lambda\in SO(1,3)_+^\uparrow) 
\eea
(generalizing \eref{itw} and \eref{itw-r}; $D(\Lambda)$ may include the 
action on a string variable), and that fulfill the completeness relation 
\bea\label{partition}
g^{MN}u_{Mn_1}(p)\ol{u_{Nn_2}(p)} = \delta_{n_1n_2}
\eea
with a suitable ``metric'' $g^{MN}$. In order for 
\eref{partition} to be fulfilled, $D(\Lambda)$ may have to be a direct
sum of tensor representations (in the bosonic case), and hence $g$ a
corresponding direct sum of tensor products of $\eta$; we present
examples for such $u_{Mn}$ below (\xref{x:x} and \sref{s:proof}).  

\newpage

Inserting the partition of unity \eref{partition} into the previous
expression, we get 
$$P_\sig = -\frac12 g^{MN} \int d^3\vec x
\,\phi_M^+(x)\lrd0\lrd\sig\phi_N^-(x) = 
-\frac14 g^{MN} \int d^3\vec x \,\wick{\phi_M\lrd0\lrd\sig\phi_N}(x),$$
where $\phi_M^+(x) = \int d\mu_m(p)\,e^{ipx}\,u_{Mn}(p)\,a_{n}^*(p)$, 
and $\phi^-_M(x)$ its hermitean conjugate. 
$\phi_M(x)=\phi_M^+(x)+\phi_M^-(x)$ is a covariant field, \ploc\ or
\sloc\ according to the choice of the intertwiners. The second equality 
holds by symmetry of the Wick product and because the operator $\lrd0$ 
vanishes on the creation-creation and annihilation-annihilation parts
of the Wick product thanks to $p_{10}=p_{20}$.  

The last expression is the desired local representation of \eref{P}.
The integrand 
\bea\label{tT}
\wt T_{\rho\sig}(x) = -\frac14 g^{MN} \,\wick{\phi_M\lrd\rho\lrd\sig\phi_N}(x)
\eea
is a first candidate for a \set, that by construction produces the
correct generators $P_\sig$ of translations. Of course, this
construction is only unique up to terms that vanish upon the 
$\vec x$-integration, and we shall see presently that we need to add
such terms in order to produce also the correct Lorentz generators.

We obtain the global form of the Lorentz generators from the 
transformation law of the creation operators:
$$U(\Lambda)a_n^*(p)U(\Lambda)^*= a_m^*(\Lambda p)
d_{mn}(W_{\Lambda,p}) \qquad
(\Lambda\in \SO(1,3)_+^\uparrow).$$
Infinitesimally: 
$$i[M_{\sig\tau},a_n^*(p)] =
\big(\delta_{nn'}\, (p_\sig\partial_{p^\tau} -
p_\tau \partial_{p^\sig})+d(\omega_{\sig\tau})^t_{nn'})a_{n'}^{*}(p),$$
where $\omega_{\sig\tau}$ is the infinitesimal Wigner ``rotation''. The
latter depends on the choice of the standard ``boosts''; we do not
display it because it is going to cancel anyway. 
What matters is that
$d(\omega_{\sig\tau})$ is anti-hermitean on $\HH_d$ because the
representation $d$ is unitary, hence
$-i(\eins_d\,p\wedge\partial_p+d(\omega)^t)_{\sig\tau}\equiv 
-i\big(\eins_d\,(p_\sig\partial_{p^\tau} - 
p_\tau \partial_{p^\sig})+d(\omega_{\sig\tau})^t\big)$ is selfadjoint 
on the one-particle space. It follows:  
\begin{lemma}\label{l:M}
The selfadjoint infinitesimal generators $M_{\sig\tau}$ of the
Lorentz transformations are the second quantization 
$$M_{\sig\tau} = -i\int d\mu_m(p)
\big((\delta_{nn'}\,p\wedge\partial_p+d(\omega)^t_{nn'})_{\sig\tau}
a^*_{n'}(p)\big) a_{n}(p)$$
of the operators
$-i(\eins_d\,p\wedge\partial_p+d(\omega)^t)_{\sig\tau}$. 
\end{lemma}
We now proceed as before with the momentum generators, inserting the
partition of unity for the momenta via an $\vec x$-integration, and
the partition of unity for the spin components via the sum
\eref{partition} over intertwiners. By partial integration in $p_1$,
the operators $-i(\eins_d\,p_1\wedge\partial_{p_1}+d(\omega)^t)$
acting on the creation operators $a^*(p_1)$ are shifted to the wave
function $u(p_1)e^{ip_1x}$ where they act like
$$-i(-p_1\wedge\partial_{p_1}+d(\omega))\big(u(p_1)e^{ip_1x}\big) =
\big((i\,p_1\wedge\partial_{p_1}-i\,d(\omega) +x\wedge p_1)u(p_1)\big)e^{ip_1x}.$$
The term $x\wedge p_1$ is treated exactly as before, and gives the
expected contribution
$\int d^3\vec x \, (x_\sig \wt T_{0\tau} - x_\tau \wt T_{0\sig})$
to $M_{\sig\tau}$. We are now going to compute the remaining term. 

The infinitesimal version of the intertwining property \eref{itw-gen}
is 
$$(p\wedge \partial_p)u(p) = (D(\Omega)+d(\omega))u(p),$$ 
therefore $i(p\wedge\partial_p-d(\omega))_{\sigma\tau}u = iD(\Omega_{\sig\tau})u$
involves the infinitesimal Lorentz transformation
$D(\Omega_{\sig\tau})$ of the intertwiner (which is the same as that 
of the field). Thus, 
$$M_{\sig\tau} = \int d^3\vec x \, (x_\sig \wt T_{0\tau} - x_\tau \wt
T_{0\sig}) + \Delta M_{\sig\tau}$$ 
where
\bea\label{DelM}
\Delta M_{\sig\tau} =  -\frac12  g^{MN} \int d^3\vec x \, 
\wick{(D(\Omega_{\sig\tau})\phi)_M\lrd0\phi_N}.
\eea
We now use Lemma~B.1 in \cite{MRS1}: For a symmetric and conserved
tensor $\Theta_{\rho\sig} = \partial^\mu[X_{\rho\mu}\lrd\sig Y
+X_{\sig\mu}\lrd\rho Y]$, where $X_{\rho\mu}$ is anti-symmetric and
both $X(x)$ and $Y(x)$ satisfy the Klein-Gordon equation, one has
$$\int d^3\vec x\, \big(x_\sig\Theta_{0\tau}-x_\tau\Theta_{0\sig}\big)
= 2 \int d^3\vec x\, X_{\sig\tau} \lrd0 Y. $$  
\begin{coro}\label{c:Tgen} Under the assumption \eref{partition}, the
  total \set\ $\wt T_{\rho\sig} + \partial^\mu\Delta T_{\rho\sig;\mu}$
  produces the correct generators of the Poincar\'e group \eref{PM},
  where $\wt T_{\rho\sig}$ is given by \eref{tT} and  
\bea\label{DelT}
\Delta T_{\rho\sig;\mu} = -\frac14 
g^{MN}\Big[D(\Omega_{\rho\mu})_{M}{}^{M'}\wick{\phi_{M'}\lrd\sig\phi_N}(x)
+ (\rho\lra\sig)\Big].
\eea
Notice that $\wt T_{\rho\sig}$ and $\partial^\mu\Delta
_{\rho\sig;\mu}$ are separately manifestly symmetric and conserved.
\end{coro}
\begin{example}\label{x:x}
For the massive Proca field of spin $s$, the standard intertwiner
$u_n^{{\rm P}(s)}(p) = (B_pE_3)^{\otimes s} T_n$ as in the proof of
\pref{p:asc} fulfills \eref{partition} with the metric 
$g = (-1)^r\eta^{\otimes r}$. The sign is due to the anti-isometric
property of the embedding $E_3:\RR^3\to\RR^4$. The field 
$A^{{\rm P}(s)}$ transforms in the representation
$D(\Lambda)=\Lambda^{\otimes s}$ of the Lorentz group. With
$(\Omega_{\sig\tau})_\mu{}^{\mu'} = \eta_{\tau\mu}\delta_\sig^{\mu'}-
\eta_{\sig\mu}\delta_\tau^{\mu'}$,
one gets 
$$\Delta M_{\sig\tau} = -(-1)^r r \int d^3x\,  \wick{A^{{\rm
      P}(s)}_{\sig\mu_2\dots\mu_s}\lrd0 A^{{\rm
      P}(s)\mu_2\dots\mu_s}_\tau}(x)$$ 
and
$$\Delta T^{{\rm P}(s)}_{\rho\sig;\mu} = -(-1)^r\frac r2  
\Big[\wick{A^{{\rm P}(s)}_{\rho\mu_2\dots\mu_s}\lrd\sig 
A^{{\rm P}(s)\mu_2\dots\mu_s}_\mu}(x) + (\rho\lra\sig)\Big].$$
This gives the reduced \set\ \eref{Tred}, found in \cite{MRS1} by a
less systematic approach. Its main part $\wt T^{{\rm P}(s)}_{\rho\sig}$ 
(without the derivative term) appeared already in Fierz' paper \cite{Fz}. 
For the same Wigner representation $(m,s)$, none of the \sloc\
intertwiners $u^{(s,r)}$ fulfills \eref{partition} separately. They
must be combined, in a manner similar to the infinite-spin case
below. In this case, \cref{c:Tgen} gives the regular \set\ \eref{Treg}.
\end{example}

\subsection{Proof of \pref{p:set}}
\label{s:proof}

In order to prove \pref{p:set}, we apply the prescription of
the preceding subsection. By \cref{c:Tgen}, we need to fulfill
\eref{partition} with intertwiners of \sloc\ fields. For the
infinite-spin representations, the representation space $\HH_d$ is
$L^2(\kappa S^1)$ with $d\mu_\kappa(\vec k)=\frac{d\varphi}{2\pi}$,
hence $\delta_{n_1n_2}$ is replaced by $\delta_\kappa(\vec k_1,\vec
k_2) = 2\pi\cdot\delta_{2\pi}(\varphi_1-\varphi_2)$. 

\begin{lemma}\label{l:partition} The completeness relation
  \eref{partition} is fulfilled by 
\bea\label{partinfspin}
\sum\nolimits_{r\geq0} (-1)^r \eta^{\mu_1\nu_1}\dots
\eta^{\mu_r\nu_r}U^{\kappa(r)}_{\mu_1\dots\mu_r}(e,p)(\vec k_1)
\ol{U^{\kappa(r)}_{\nu_1\dots\nu_r}(e,p)(\vec k_2)} 
= \delta_\kappa(\vec k_1,\vec k_2).\quad
\eea
\end{lemma}
Proof: By \pref{p:Ur}(ii), the left-hand side equals 
$u^{\kappa(0)}(e,p)(\vec k_1)\ol{u^{\kappa(0)}(e,p)(\vec k_2)}$ times 
$$\big(\sum\nolimits_{r\in\ZZ}
e^{-ir(\varphi_1-\varphi_2)}\big) =
2\pi\delta_{2\pi}(\varphi_1-\varphi_2)=\delta_\kappa(\vec k_1,\vec
k_2),$$
and at $\vec k_1=\vec k_2$ the factor is $=1$ because
$u^{\kappa(0)}(e,p)(\vec k)$ is a complex phase. \qed 
\begin{remark}\label{r:ee}
One sees why it is crucial that the two strings in \eref{partinfspin} 
are equal: otherwise the phase factors $u^{\kappa(0)}\ol{u^{\kappa(0)}}$ 
would fail to cancel. It is also essential that $u^{\kappa(0)}$ is a
function (\pref{p:k-bd}), since otherwise the product of distributions
involving  $\frac {e}{(ep)_+}$ and $\ol{\frac {e}{(ep)_+}}=\frac
{e}{(ep)_-}$ were ill-defined.
Because the factors $u^{\kappa(0)}$ only appear through the inductive limit
\aref{a:ind}, cf.\ \rref{r:conv}, they are absent in the case of finite
spin, and the analogous requirement of coinciding strings may be
dropped, as in \eref{Treg}. 
\end{remark}
We can thus apply \cref{c:Tgen} {\em mutatis mutandis}. Apart from the
specific partition of unity \eref{partinfspin}, the only change is the
dependence of the intertwiners on $e$, which are also transformed
along with the Lorentz tensors by $D(\Lambda)$, specifying
\eref{itw-gen} as 
$$U^{\kappa(r)}(e,\Lambda p) = 
(\Lambda^{\otimes r}\otimes d_\kappa(W_{\Lambda,p}))U^{\kappa(r)}(\Lambda\inv e,p) 
\qquad (\Lambda\in \SO(1,3)_+^\uparrow). $$
Therefore, $D(\Omega_{\sig\tau})$ contains, besides the infinitesimal
Lorentz matrices, the additional term $-(e\wedge
\partial_e)_{\sig\tau}$, and $\Delta T$ is a sum
of two terms: 
\bea\label{DelTinf}
\Delta T^{\kappa}_{\rho\sig;\mu}(e,x) = \Delta_1
T^{\kappa}_{\rho\sig;\mu}(e,x) + \Delta_2 T^{\kappa}_{\rho\sig;\mu}(e,x)
\eea
where
$$\Delta_1 T^{\kappa}_{\rho\sig;\mu}(e,x) = - \sum\nolimits_{r\geq0}
(-1)^r\frac r2\Big[\wick{\Phi^{\kappa(r)}_{\rho\mu_2\dots\mu_r}(e)\lrd\sig 
\Phi^{\kappa(r)\mu_2\dots\mu_r}_\mu(e)}(x)  
+(\rho\lra\sig)\Big]$$
similar as in \xref{x:x}, and 
$$\Delta_2 T^{\kappa}_{\rho\sig;\mu}(e,x) =-
\sum\nolimits_{r\geq0}\frac{(-1)^r}8
\Big[\wick{\Phi^{\kappa(r)}_{\mu_1\dots\mu_r}(e)
(e\wedge\lrd e)_{\rho\mu}\lrd\sig\Phi^{\kappa(r)\mu_1\dots\mu_r}(e)}(x)
+ (\rho\lra\sig)\Big].$$
With the computation of $\wt T$, coinciding with the expression
displayed in \eref{Tinf}, and the specification of 
$\Delta T = \Delta_1 T +\Delta_2 T$, the proof of \pref{p:set} is complete.
\qed
\begin{remark}\label{r:mix}
Derivatives w.r.t.\ $e$ seem to mix
$\Phi^{\kappa(r)}$ with $\Phi^{\kappa(r\pm 1)}$ by \eref{coup}, but because
\eref{coup} involves a symmetrization, we cannot simply write
$(e\wedge\partial_e)\Phi^{\kappa(r)}$ as a combination of 
$\Phi^{\kappa(r\pm 1)}$. Recall, however, that the fields
$\Phi^{\kappa(r)}(e)$ are simultaneously defined for all $e$ on the
same Hilbert space, and the derivative w.r.t.\ $e$ does not change 
the localization. Therefore, each $T^{\kappa(r)}$ in \eref{Tinf} 
is well-defined. 
\end{remark}
The potential problems due to the infinite summation over $r$ in
\eref{Tinf} will be discussed in the next section. 

The same general strategy outlined in \sref{s:qsets} applies to
conserved currents of complex fields. In the case of infinite spin,
the partition of unity \eref{partinfspin} inserted into the charge operator
$$Q = \int d\mu_0(p) d\mu_\kappa(\vec k)
\big(a^*(p,\vec k)a(p,\vec k)-b^*(p,\vec k)b(p,\vec k)\big)$$
gives rise to the current 
\bea\label{curr}
J^{\kappa}_\rho(e,x) = i\sum\nolimits_{r\geq0} (-1)^r
\wick{\Phi^{\kappa(r)*}_{\mu_1\dots\mu_r}(e)\lrd\rho
  \Phi^{\kappa(r)\mu_1\dots\mu_r}(e)}(x).
\eea
\begin{remark}\label{r:subalg}
Infinite-spin fields admit no subalgebra of compactly localized
observables (``field strengths'' or currents) whose charged sectors
they would generate from the vacuum \cite{K,LMR}. Therefore,
``neutral'' operators like the current densities or \set\ cannot be
\ploc\ as in the massive case; the localization on a pair of opposite
strings seems to be the best that is possible.
\end{remark}

\subsection{Properties of the infinite-spin \set}
\label{s:props}
We present here some qualitative material that helps to assess the 
mathematical nature of fields like the infinite-spin \set\ \eref{Tinf}. 
The rigorous analytical treatment is beyond the scope of this article. 

\subsubsection{Matrix elements}
\label{s:matrix}

For a generic one-particle state
$\Psi =\int d\mu_0(p)\,d\mu_\kappa(\vec k)\, \psi(p,\vec k) \, a^{*}(p,\vec
k)\Omega$, we compute matrix elements 
$$(\Psi,\Phi^{\kappa(r)}(e,x)\Omega)= 
\int d\mu_0(p)\, e^{ipx}\, (\psi(p),U^{\kappa(r)}(e,p))_\kappa $$
where $(\cdot,\cdot)_\kappa$ is the scalar product of $L^2(\kappa
S^1)$. Let $R_{e,p}\bpm\cos\alpha_{e,p}\\[-.5mm]\sin\alpha_{e,p}\epm$
parametrize the $1$-$2$-part of $B_p\inv q_e(p)$, so that
$u^{\kappa(0)}(e,p)(\vec k)=e^{-i(q_eE_p(\vec k))} = e^{i\kappa
  R_{e,p}\cos(\varphi-\alpha_{e,p})}$. 

For $\psi^n(p,\vec k) = \psi(p)e^{in\varphi}$, the $\vec k$-integration
produces Bessel functions \eref{bessnu}.
Thus, with $\Eps_\pm(e,p) =J_e(p)E_p\eps_\pm$ as in \pref{p:Ur}
$$ (\psi^n(p),U^{\kappa(r)}(e,p))_\kappa =
(-1)^r\ol{\psi(p)}\omega^\kappa(e,p)\sum\nolimits_{\pm}
J_{n\pm r}(\kappa R_{e,p}) e^{-i(n\pm r)(\alpha_{e,p}-\frac\pi2)} 
\cdot \Eps_\pm(e,p)^{\otimes r}.$$
For $r=0$, the sum of two terms is replaced by 
$J_{n}(\kappa R_{e,p}) e^{-in(\alpha_{e,p}-\frac\pi2)}$. 

\newpage

This formula is not particularly useful, but it shows that there is no
correlation between $r$ and $n$ (cf.\ \rref{r:sharp}), and that
infinite sums over $r$, as in the \set\ or the current, are
potentially dangerous, as already pointed out in \rref{r:warning}. 

Let us exemplarily investigate this issue in various situations: matrix
elements, \tpf, and commutators of the \set\ or the current. In order
to simplify the presentation, we consider the scalar Wick square
\bea\label{W}
W^{\kappa}(e,x) &=& \sum\nolimits_{r=0}^\infty
(-1)^r \wick{\Phi^{\kappa(r)\mu_1\dots\mu_r}(e) 
\Phi^{\kappa(r)}_{\mu_1\dots\mu_r}(e)}(x),
\eea
in which the characteristic features of the infinite sum can
be seen as well. For the actual \set, one basically has to insert
polynomial factors of $p$ corresponding to the derivatives 
$\lrd\rho\lrd\sig$, and add a similar contribution from $\Delta T$.  

We compute matrix elements of the Wick square, for simplicity in
one-particle states $\Psi_i$ with wave functions $\psi^0_i(p)$ (i.e., $n=0$):
\bea\notag 
(\Psi^0_1,W^{\kappa}(e,x)\Psi^0_2) = 2\int d\mu_0(p_1)d\mu_0(p_2)\,
e^{i(p_1-p_2)x}\, \ol{\psi_1(p_1)}\psi_2(p_2)\cdot
\omega^\kappa(e,p_1)\ol{\omega^\kappa(e,p_2)}\cdot \\ \notag\cdot
\sum\nolimits_{\nu\in\ZZ}
e^{-i\nu(\alpha_{e,p_1}-\alpha_{e,p_2})} 
J_{\nu}(\kappa R_{e,p_1})J_{\nu}(\kappa R_{e,p_2}).\hspace{30mm}
\eea
Similar expressions with $J_{\nu+n_1}J_{\nu+n_2}$ hold for matrix
elements in states with $n_i\neq0$, or for matrix elements between the
vacuum and two-particle states. 

The point is that the sum over $r\in\ZZ$ is absolutely convergent
thanks to the Cauchy-Schwartz inequality applied to the
square-summability of the Bessel functions: 
\bea\label{JJ}
\sum\nolimits_{\nu\in \ZZ} J_\nu(x)^2=1.
\eea
Together with Wick's theorem for matrix elements between
multi-particle states, this observation supports our 
\begin{conj}\label{cj:matrix} 
  The Wick square \eref{W} and likewise the \set\ \eref{Tinf} and the
  current \eref{curr} have finite matrix elements in states of finite
  particle number and finite energy. Because such states are dense in
  the Fock space, these fields exist as quadratic forms with a dense domain. 
\end{conj}

\subsubsection{Two-point functions and vacuum fluctuations}
\label{s:tpf}

The \tpf\ of $W^{(\kappa)}$ is a double sum over $r$ and 
$r'$ of the fully contracted squares of \tpf s
$(\Omega,\Phi^{\kappa(r)}_{\ul\mu}(x)\Phi^{\kappa(r')}_{\ul\nu}(y)\Omega)$
given in \pref{p:Phirr'}.

If $R_{e,e',p}\bpm\cos\alpha_{e,e',p}\\[-.5mm]\sin\alpha_{e,e',p}\epm$
parametrizes the $1$-$2$-part of $B_p\inv (q_e(p)-q_{e'}(p))$, then 
\bea\notag
(\Omega,W^\kappa(e,x)W^\kappa(e',x')\Omega) = 2
\int d\mu_0(p_1)d\mu_0(p_2)\,e^{-i(p_1+p_2)(x-x')}
\prod_{i=1,2} \ol{\omega^\kappa(e,p_i)}\omega^\kappa(e',p_i)\cdot \\ \notag
\cdot\sum\nolimits_{\nu,\nu'\in\ZZ} (-1)^{\nu-\nu'}
e^{i(\nu+\nu')(\alpha_{e,e',p_1}-\alpha_{e,e',p_2})} J_{\nu+\nu'}(\kappa
R_{e,e',p_1})J_{\nu+\nu'}(\kappa R_{e,e',p_2}).\hspace{10mm}
\eea
The problem is that the double sum may not exist, because the 
square-summability and Cauchy-Schwartz argument (as for the matrix
elements) does not apply: the convolution
product of square-summable sequences need not be square-summable.

Of course, smearing with test functions does not help. This supports our 
\begin{conj}\label{cj:tpf} The \tpf s of $W^{\kappa}$, $T^{\kappa}_{\rho\sig}$,
and  $J_\rho^{\kappa}$ do not exist.    
\end{conj}
Mathematically, this means that the \set\ does not exist as an
operator-valued distribution with a stable domain containing the
vacuum vector, as required by the Wightman axioms.
In physical terms, the divergence of the \tpf\ signals infinitely
strong vacuum fluctuations. Stress-energy tensors that exist as quadratic 
forms (\cjref{cj:matrix}), but not as Wightman fields (\cjref{cj:tpf}),
occur also for generalized free fields \cite{DR}. Here, the vacuum
fluctations are also divergent, but not because of the infinitely
degenerate spin component, but because a continuous mass
distribution cannot be ``square-summable''. 

\subsubsection{Commutators}
\label{s:Comm}

We have seen that the decisive difference between the ``good''
behaviour of matrix elements and the ``bad'' behaviour of \tpf s is
due to the summation structure. Let us therefore study the 
commutator of the Wick square with a field just under this aspect.

The summation structure of the commutator is
the same as that of a matrix element
$(\Psi,W^{\kappa}\Phi^{\kappa(r)}\Omega)$ with a one-particle
state. Choose for simplicity $r=0$, and $\Psi^n$ of helicity $n$  as in
\sref{s:matrix}. Then, one computes 
\bea\notag 
(\Psi^n,W^\kappa(e,x)\Phi^{\kappa(0)}(e',x')\Omega) = \hspace{80mm}
\\ \notag \hspace{10mm} =2\int
d\mu_0(p_1)d\mu_0(p_2) \, e^{ip_1x}e^{-ip_2(x-x')}\cdot
\ol{\psi^n(p_1)}\cdot
\omega^\kappa(e,p_1)\ol{\omega^\kappa(e,p_2)}\omega^\kappa(e',p_2)\cdot
\\ \notag \cdot\sum\nolimits_{\nu\in\ZZ}
i^ne^{-i(n+\nu)\alpha_{e,p_1}}e^{i\nu\alpha_{ee',p_2}} J_{\nu+n}(\kappa
R_{e,p_1})J_\nu(\kappa R_{e,e',p_2}). \hspace{10mm}
\eea
This sum is absolutely convergent, as for the matrix elements
above. The same expression with a different $i\eps$ prescription
(hidden in the argument $R_{e,e',p_2}$ of the Bessel function) holds
for the matrix element $(\Psi,\Phi^{\kappa(r)}W^{\kappa}\Omega)$, and hence
the sum also converges for the commutator. This sketch of an argument
supports our  
\begin{conj}\label{cj:comm} The commutators of $W^{\kappa}$, 
$T^{\kappa}_{\rho\sig}$, and $J_\rho^{\kappa}$ with the linear fields
$\Phi^{\kappa(r)}_{\mu_1\dots\mu_r}$ exist and can be defined as
derivations on the algebra generated by smeared fields.    
\end{conj}
In view of \cjref{cj:tpf}, this property would rescue the \set\ as a
``good'' physical quantity. Namely, the prime role of
the \set\ in quantum field theory is to generate infinitesimal
Poincar\'e transformations via commutators. Of course, other technical
issues remain concerning the convergence of the commutator with a
smeared \set\ when the smearing functions becomes constant in space
and sharp in time. 

More interestingly, \cjref{cj:comm} could also secure the existence of the
perturbative expansion of a coupling of infinite-spin matter to
linearized gravity via its \set, because this expansion is a series in
retarded commutators. 

More detailed investigations of these issues are beyond the scope of
this paper. 

\subsubsection{Thermal states: equation of state and equipartition}
\label{s:KMS}

Further interesting quantities to study are the energy density and the
pressure in thermal equilibrium. 

The computation of thermal expectation values of quadratic fields
$\wick{XY}(x)$ is most easily done by first considering
$\omega_\beta(X(x)Y(x'))$ at $x\neq x'$ and using the KMS condition in 
momentum space (e.g., \cite[Eq.~(16)]{MSY2}). It determines the
thermal \tpk\ on the negative mass shell by ``detailed balance'':  
$$\merw{m,\beta}{X,Y}(-p) = e^{-\beta p^0}\cdot\merw{m,\beta}{Y,X}(p).$$
Then one exploits the fact that the commutator is the same
in the vacuum and in the thermal state, hence $\merw{m,{\rm
    vac}}{X,Y}(p) = \merw{m,\beta}{X,Y}(p)
-\merw{m,\beta}{Y,X}(-p)$. 
This implies
$$\merw{m,\beta}{X,Y}(p) = \frac 1{1-e^{-\beta p^0}}\cdot
\merw{m,{\rm vac}}{X,Y}(p)$$
on the positive mass shell. Subtracting the vacuum expectation value,
one gets
$$\omega_\beta(\wick{X(x)Y(x')}) = \int d\mu_m(p)\frac1{e^{\beta  p^0}-1}
\Big[\merw{m,{\rm vac}}{X,Y}(p)e^{-ip(x-y)}
+\merw{m,{\rm vac}}{Y,X}(p)e^{ip(x-y)}\Big].$$
Here, one can put $x=x'$, and thus obtains the thermal expectation value
of $\wick{XY}(x)$ from the vacuum \tpk s.

This very efficient method reduces the computations of thermal
expectation values to the inspection of the vacuum kernels, without
any computation of partition functions in finite volume. It
immediately gives
the thermal energy density $\eps=\omega_\beta(T^{{\rm red}(s)}_{00})$ and the
pressure $p=\omega_\beta(T^{{\rm red}(s)}_{ii})$ of massive matter of
finite spin 
\bea\label{eps}
\eps=
\frac{(2s+1)}{2\pi^2\beta^4}\cdot I_+(\beta m), \quad 
3p = \frac{(2s+1)}{2\pi^2\beta^4}\cdot I_-(\beta m), 
\eea
where $I_\pm(x)=\int_0^\infty \frac {u^3\,du}{e^{\sqrt{u^2+x^2}}-1}
\big(\frac{\sqrt{u^2+x^2}}u\big)^{\pm1}$. 
The manifest factor $2s+1$ reflects the law of equipartition. 
The result is independent of the choice of the \set,
because KMS states are translation invariant, hence the derivative
terms by which various \set s differ, do not contribute. 
Interestingly, the individual contributions from
$t^{(s,r)}$ in \eref{Treg} depend on $e_1$ and $e_2$, while only their
sum is independent of the strings. E.g., for $s=1$, the contributions are
$1-m^2(q_{e_1}q_{e_2})$ from $r=0$ and $2+m^2(q_{e_1}q_{e_2})$ from $r=1$. In
the \pll, each contribution from $t^{(s,r)}$ converges to zero
(because of the factor $F_s(1)$ in \eref{aa=FG}), but their sum diverges
as $2s+1$ (because of \eref{eps}). The total energy density per degree of
freedom and the pressure per degree of freedom remain finite, and obey
the usual massless equation of state.

The \sloc\ \set\ $T^{(s)m=0}$ of massless fields of finite helicity 
$\vert h\vert>0$ gives the factor $2$, as expected. 
At $m=0$, the finite values $I_+(0)=I_-(0)=\frac{\pi^4}{15}$ reproduce
the Stefan-Boltzmann law and the massless equation of state
$p(\eps)=\frac13\eps$. (Interestingly, while the trace of the reduced
\set\ is non-zero and not even defined at $m=0$, its thermal 
expectation value vanishes in the limit $m\to0$). 

For the infinite-spin \set, the contribution from each $T^{\kappa(r)}$
is $2$ (resp.\ $1$ for $r=0$). Thus, the sum over $r$ diverges as
$2r+1$,  confirming the heuristic expectation. Wigner argued in
\cite{W48} that this need not imply that infinite-spin matter must be
unphysical, because it might never reach thermal equilibrium. Of
course, this question cannot be physically addressed without a
dynamical model for the coupling to ordinary matter. E.g., Schroer
\cite{S17} argues that infinite-spin matter cannot couple to ordinary
matter because there is no interaction Lagrangean that yields a
string-independent action, as is needed to preserve causality in the
quantum perturbation theory \cite{MRS1}. Thus infinite-spin matter is
``inert'', and has no mechanism to approach thermal equilibrium at all.

\section{Conclusion}

We have ``liberated quantum field theory from its classical crutches''
(in the words of P. Jordan) by finding a construction scheme for
covariant quantum stress-energy tensors that does not refer to a
classical action. The method is applicable to arbitrary (in this
paper: integer or infinite) spin. Auxiliary fields implementing
higher-spin constraints, negative probability states, and compensating
ghosts never appear.  

Instead, the prescription is based on Wigner's unitary representation
theory of the Poincar\'e group and Weinberg's construction
of covariant quantum fields with the help of intertwiners whose
analytic properties entail the localization properties of the fields. 

The achieved \set s are not unique, depending on a choice of
intertwiners fulfilling the localizing completeness relation
\eref{partition}. However, their densities all differ by ``irrelevant
derivatives'' in the sense that they all produce the same Poincar\'e
generators when integrated over space at a fixed time. Even for low
spin, our ``reduced'' \set s (\xref{x:x}) differ from the canonical or
Hilbert \set s by irrelevant derivative terms.  

We applied this method in the case of the infinite-spin representations,
where the best possible localization is on strings of the form
$S_e(x) = x+\RR_+\cdot e$. In this case, the completeness relation
\eref{partition} requires an infinite direct sum of representations of
the Lorentz group, which causes the \set\ to be an infinite sum of
quadratic expressions in the corresponding \sloc\ fields. We sketched in
\sref{s:props} the ensuing analytical implications (problems of
convergence) with indications for ``one bad and two good'' features.  

The good features are that matrix elements and commutators of the
\set\ are well-behaved, while its correlations functions suffer from
infinite vacuum fluctuations (the price of infinite spin).

The involved \sloc\ fields are defined on the Fock space over Wigner's
infinite-spin representation. We
constructed these fields as \pll s of tensor fields of increasing spin
and decreasing mass with fixed Pauli-Lubanski parameter
$\kappa^2=m^2s(s+1)$. Although it is not needed for the
determination of the \set, this approximation is of some interest
of its own. E.g., it exhibits how the dynamical coupling between
escort fields $A^{(s,r)}$ of different $r$, that goes to zero with the
mass at fixed $s$, remains stable (proportional to $\kappa$) when the
spin increases. This may also play a role in higher spin theories. 

\bigskip

{\bf Acknowledgements.}
I am grateful for invitations to the Universidade de Juiz de Fora,
where this project has started, and to the University of York, where 
parts of it have been done. I thank Bert Schroer and Jens Mund for
helpful discussions and Jakob Yngvason for his encouraging
interest. This work had been impossible without their groundbreaking
work on the infinite-spin representations. I thank the referee for
pointing out Ref.~\cite{McK}.

\appendix

\section{\pll\ of Wigner representations}
\label{a:ind}
\setcounter{equation}{0}
We give a non-techical presentation of the \pll\ of the (one-particle)
Wigner representations. For a more rigorous treatment, see \cite{McK}. 

The standard reference vector of the massive Wigner representation
$(m,s)$ is $p_m=(m,0,0,0)$. Its stabilizer group is $\Stab(p_m)=
\SO(3)\subset \SO(1,3)$. We denote its generators $L_i$ as usual, and
$K_i$ the generators of the boosts. The reference
vector for a massless Wigner representation is $p_0=(1,0,0,1)^t$. We
approximate it by massive vectors $p_\tau=m(\cosh\tau,0,0,\sinh\tau)^t$ with
$e^\tau=\frac 2m$ in the limit $m\to0$. Let $B^\tau$ be the Lorentz
$3$-boost such that $B^\tau p_m=p_\tau$. The stabilizer group of
$p_\tau$ is $\Stab(p_\tau) = B^\tau \SO(3) B^\tau{}\inv$. 

For the generators of $\Stab(p_\tau)$ one computes $L_3^\tau:=B^\tau L_3
B^\tau{}\inv = L_3$ and $L_i^\tau := B^\tau L_i B^\tau{}\inv = \cosh\tau L_i +
\sinh\tau K_i$ for $i=1,2$. Thus, in $\so(1,3)$
$$\lim_{\tau\to\infty} 2e^{-\tau} L^\tau_i = Q_i$$
are the ``translation'' generators of $\Stab(p_0)=E(2)$, while $L_3$ is the
generator of the rotations in $E(2)$. 

Now, we choose the unitary representation of $SO(3)$ with standard
orthonormal basis $\ket{s,n}\in \HH_s$ for each spin, for which
$L_3\ket{s,n}=n\cdot\ket{s,n}$ 
and $L_\pm \ket{s,n}=\sqrt{s(s+1)-n(n\pm1)}\cdot\ket{s,n\pm 1}$. For
increasing $s$, we isometrically embed $\HH_s\to\HH_{s+1}$ by
$\ket{s,n}\mapsto \ket{s+1,n}$. The \pll\ is the
inductive limit of the representations of the generators $2e^{-\tau}
L^\tau_i$ and $L^\tau_3$ in this sequence of representations as
$s\to\infty$, while $m^2s(s+1)=\kappa^2$ is constant. Thus,
$2e^{-\tau} = m = \kappa(s(s+1))^{-\frac12}$, hence 
$$2e^{-\tau} L_\pm^\tau\ket{s,n} = \kappa 
\Big(1-\frac{n(n\pm1)}{s(s+1)}\Big)^{\frac12}\cdot\ket{s,n\pm1}, 
\quad L_3^\tau\ket{s,n}=n\cdot\ket{s,n}.$$ 
In the inductive limit, this becomes the representation $D_{\kappa}$ of $E(2)$:
$$Q_\pm\ket{n} = \kappa \cdot \ket{n\pm1}, \quad L_3\ket{n}=n\cdot\ket{n}.$$ 
(Representing $\ket n\in \HH_\kappa = L^2(\kappa S^1)$ by the wave
function $\psi_n(\varphi)=e^{in\varphi}$, $Q_\pm$ act by
multiplication with $\kappa e^{\pm i\varphi}$, hence $\vec Q$ act by
multiplication with $\vec k=\kappa(\cos\varphi,\sin\varphi)$.) 

Once the representation $d$ of the respective stabilizer group is
specified, the corresponding induced Wigner representation of the
Poincar\'e group is defined on $L^2(H_m,\HH_d)$. The translations act on
wavefunctions $\psi(p)$ with values in $\HH_d$ by multiplication with
$e^{ipx}$, and the Lorentz transformations act by 
$$(U(\Lambda)\psi)(p) = d(W_{\Lambda,\Lambda\inv p})\psi(\Lambda\inv p),$$
where $W_{\Lambda,p}=B_{\Lambda p}\inv\Lambda B_p$ is the Wigner
``rotation'' in the stabilizer group of the respective reference
vector $p_0$. It depends on the choice of the standard ``boosts''
$B_p$ that take $p_0$ to $p$, but the dependence is a unitary
equivalence of $U$. This unitary equivalence acts on $L^2(H_m,\HH_d)$
as a multiplication with a function $H_m\to U(\HH)$, and is of course
irrelevant for abstract properties. 

The inductive limit of representations of the stabilizer groups,
outlined before, naturally extends to the induced representations of the
Poincar\'e group.


\begin{thebibliography}{99}
\bibitem{G17} C. Ga\ss\ has suggested this simpler proof of a stronger
  statement than in a previous version.
\end{thebibliography}
\end{document}